\documentclass[a4paper,11pt]{article}
\pdfoutput=1 

\usepackage{jcappub} 

\usepackage{newtxtext,newtxmath}

\usepackage[T1]{fontenc} 

\let\vec\mathbf
\let\sun\odot

\DeclareRobustCommand{\VAN}[3]{#2}
\let\VANthebibliography\thebibliography
\def\thebibliography{\DeclareRobustCommand{\VAN}[3]{##3}\VANthebibliography}

\usepackage{graphicx}	
\usepackage{subcaption}
\usepackage{amsmath}	
\usepackage{hyperref}

\title{A machine learning-based method for populating dark matter halos in N-body simulations with substructure}

\author[a,b]{Milad Noorikuhani,}
\author[a]{Jeremy Tinker}

\affiliation[a]{Center for Cosmology and Particle Physics, Department of Physics, New York University,\\ 726 Broadway, New York, NY, 10003, USA}
\affiliation[b]{Department of Physics, Sharif University of Technology, P.O.Box 11155-9161, Tehran, Iran}

\emailAdd{mn2317@nyu.edu}
\emailAdd{jeremy.tinker@nyu.edu}

\abstract{Dark matter N-body simulations that resolve halo substructure (subhalos) with high accuracy can be used for generating reliable catalogs of central and satellite galaxies via galaxy-halo connection models. However, such simulations are computationally expensive, especially when large numbers of realizations are required for statistical analyses such as robust estimations of covariance matrices for (statistical) cosmological quantities. In this work, we present a fast method for populating dark matter halos in a given simulation box with subhalos extracted from a high-resolution simulation box with the same cosmology but arbitrary initial conditions. For each halo in a given test set, the method first predicts whether it hosts at least one subhalo above a given mass threshold using a classifier built on a decision tree regressor. Then, for each predicted host, the method finds another host halo in a given high-resolution box, using a nearest neighbor search in the space of selected halo properties, and appropriately maps subhalos from that halo to the test one. By applying the method to different test sets, we make predictions for abundances, distributions and three-dimensional and projected (two-dimensional) two-point correlation functions of various subhalo populations and sub-populations. In most cases, the percent errors of our predictions are well below $5\%$. In general, the method can be used to populate halos in low-resolution simulation boxes with subhalos from a high-resolution simulation and also to investigate which halo properties are (more strongly) correlated with subhalo abundance and clustering in a given dark matter simulation.}


\begin{document}
	\maketitle
	\flushbottom

\section{Introduction}

Modern galaxy surveys, such as Euclid \cite{2025A&A...697A...1E}, DESI \cite{2016arXiv161100036D} and the Vera C. Rubin Observatory LSST  \cite{2019ApJ...873..111I}, provide the community with unprecedented opportunities for studying the abundance and clustering of galaxies with high precision thanks to their wide sky coverage and depth. For maximal extraction of information from such surveys, one needs  to build a bridge between galaxy observable quantities and dark matter halo properties, which can be done through models of galaxy-halo connection (see e.g. \cite{2018ARA&A..56..435W} for a review). Subhalo abundance matching (SHAM) is one of such models where one or more central/satellite galaxy properties (e.g. stellar mass, luminosity, etc.) are matched to one or more halo/subhalo properties (e.g. mass, maximum circular velocity, etc.) via different techniques \cite{2004ApJ...609...35K, 2004MNRAS.353..189V,  2006MNRAS.371.1173V, 2004ApJ...614..533T, 2006ApJ...647..201C, 2010ApJ...717..379B, 2013ApJ...771...30R, 2013MNRAS.436.2286M, 2013MNRAS.435.1313H, 2022ApJ...940...13D, 2023MNRAS.520..489C}.

To mention some applications, different versions of abundance matching have been used for studying the relation between galaxy stellar mass and halo mass (e.g. \cite{2010ApJ...710..903M, 2010MNRAS.404.1111G, 2010MNRAS.402.1796W, 2010ApJ...717..379B}), modeling star formation and growth histories of galaxies (e.g. \cite{2009ApJ...696..620C, 2013MNRAS.428.3121M, 2013ApJ...770...57B}), investigating assembly bias (e.g. \cite{2016MNRAS.460.1457S, 2017ApJ...834...37L}), creating mock galaxy catalogs from dark matter N-body simulations that are used for various galaxy-halo connection and cosmological analyses \citep[e.g.][]{2021MNRAS.505..325G, 2023MNRAS.524.2489C, 2023MNRAS.519.4253L, 2023ApJ...954..131B, 2023MNRAS.525.4257C, 2024MNRAS.527.6950Y, 2024A&A...689A..66O, 2025A&A...697A.226O, 2026JCAP...05..002F} and constraining cosmological parameters from galaxy clustering and galaxy-galaxy lensing signals (e.g. \cite{2026MNRAS.545f2143M}).

Considering halos and subhalos with masses above a given threshold in a realization of the universe, subhalos are usually far less abundant compared to halos (partly because they reside in more massive halos which are less abundant). Consequently, for most galaxy types, satellites constitute a minority of the corresponding galaxy population compared to centrals. Nonetheless, it is crucial to have reliable models for phase space distribution and other properties of subhalos at a given epoch as they can improve the overall performances of the SHAM techniques in all of their aforementioned applications, especially for galaxy populations with larger satellite fractions. In addition, galaxy clustering statistics at highly nonlinear scales (e.g. scales below $\sim 1 h^{-1}\text{Mpc}$) are strongly sensitive to the fraction of satellite galaxies (e.g. \cite{2013ApJ...771...30R}) which are originally hosted by subhalos. 

Previous works have studied subhalo abundance and its scatter and their dependence on host halo properties and also subhalo radial distributions, mass/velocity functions and their evolutions \citep[e.g.][]{2004ApJ...604L..73L, 2004MNRAS.355..819G, 2005ApJ...618..557N, 2005MNRAS.359.1029V, 2016MNRAS.455..158V, 2005ApJ...624..505Z, 2005ApJ...629..219Z, 2005MNRAS.364..424W, 2008MNRAS.386.2135G, 2008MNRAS.391.1685S,  2015ApJ...810...21M,10.1093/mnras/stw439, 2017MNRAS.472..657J, 10.1093/mnras/stw440, 2016MNRAS.457.1208H, 2022MNRAS.509.5305S, 2022MNRAS.509.5316S, 2022MNRAS.511..641S, 2026MNRAS.545f2099S}.  Some of such works also provide models for populating host halos with subhalos (e.g. \cite{2005ApJ...624..505Z, 2016MNRAS.457.1208H}). While these models are valuable tools for generating subhalo populations, they are usually based on simplifying assunptions about the evolution of subhalos after accretion (for example their tidal evolution) and also the spatial distribution of subhalos within their hosts which is often assumed to be spherically symmetric (on the anisotropy of subhalo distributions, see e.g. \cite{2005ApJ...629..219Z, 2005MNRAS.364..424W, 2014MNRAS.442.1197H, 2025MNRAS.538..963M}). Furthermore, to provide optimal tools for methods like SHAM, it is preferable to create subhalo populations that can be sorted according to various subhalo properties and their combinations, not just a few. This enables one to find the best matching proxy/proxies among different subhalo properties for a given galaxy type. Therefore, in general, we need tools that provide more accurate and broader picture of subhalo distributions and properties at a given epoch, for a given cosmology.

High-resolution dark matter N-body simulations that resolve subhalos with high accuracy are among such tools. However, such simulations are computationally costly and expensive computational resources are required for running a single simulation of this kind in a box whose size is relevant for cosmological analyses. The computational cost becomes prohibitive when numerous such simulations are needed in order to perform statistical analyses such as reliable estimations of covariance matrices for different (statistical) quantities that are used in cosmological studies (e.g. clustering statistics, abundances, etc.).

In recent years, numerous methods have been proposed based on different machine learning (ML) and other algorithms to generate outputs of high-fidelity dark matter simulations while circumventing their computational complexities. Broadly speaking, some of these include methods/algorithms that take as input simulation boxes with (low-resolution) linear dark matter fields or low-resolution nonlinear (evolved) fields and output high-resolution nonlinear dark matter fields or, directly, halo populations \citep[e.g.][]{2017MNRAS.465.4658M, 2019MNRAS.483L..58B, 2019PNAS..11613825H, 2019PhRvD.100d3515K, 2020MNRAS.495.4227K, 2020arXiv201200240A, 2021PNAS..11822038L, 2021MNRAS.507.1021N, 2021arXiv211106393S, 2022ApJ...930..115K,  2024MNRAS.528..281Z, 2025OJAp....8E..13Z, 2024OJAp....7E.104S, 2024PhRvD.109l3536R, 2024A&A...690A.236D, 2024arXiv240911401P, 2025PhRvD.112j3503P}.  Other methods include field level emulators and other algorithms that are trained to generate high-resolution nonlinear dark matter fields or, directly, halo populations \citep[e.g.][]{2023ApJ...952..145J, 2025JCAP...03..072J, 2024PhRvD.109l3531C, 2026PhRvD.113f3520C}. ML-based algorithms have also been used to model/investigate different aspects of galaxy-halo connection (e.g. \cite{2016MNRAS.457.1162K, 2022MNRAS.515.2733D, 2022MNRAS.514.2463D, 2022MNRAS.514.4026S, 2023mla..confE..21L, 2024ApJ...976...37W}) and to generate galaxy populations from dark matter fields or halo/subhalo populations (e.g. \cite{2013ApJ...772..147X, 2020arXiv201200186K, 2023MNRAS.518.2903I, 2024arXiv240800839B}). 

Among the ML-based works, some of them produce outputs with high enough resolutions that enable subhalos to be resolved and their statistics to be analyzed (e.g. \cite{2021MNRAS.507.1021N, 2025OJAp....8E..13Z}). Another work includes emulating subhalo/galaxy populations for milky way mass halos with varying dark matter and astrophysical parameters \citep{2026ApJ...997..336N}. There is also a work that generates subhalo populations for a specific halo mass for use in strong lensing analyses \citep{2025OJAp....8E.103L}. With regard to subhalo generation for use in cosmological (statistical) analyses, despite significant achievements, there is still room for improvement. As an example, it is beneficial (e.g. for SHAM applications) to create subhalo populations with clustering statistics that are accurate at percent level for different subhalo samples selected based on different ranges of subhalo properties.  It is also useful to have high-accuracy predictions for average subhalo abundances as a function of various host halo properties. 

In this work, we present a fast method that can be used for two main purposes. First, it can be used for creating subhalo populations in (low-resolution) dark matter simulation boxes with subhalos drawn from a high-resolution dark matter simulation box. The boxes are assumed to have the same cosmologies but their initial conditions can be different. The method makes it significantly faster to perform statistical analyses such as estimations of covariance matrices for different (statistical) quantities that are related, directly or indirectly, to subhalos by providing a fast way to create subhalo populations for multiple low-resolution boxes. Second, it can be used as a fast tool to determine which halo properties are (more) correlated with subhalo abundance and clustering in a given dark matter simulation. For halos in a simulation box, the method first predicts which of them host substructure with masses above a threshold and then populates the predicted hosts with subhalos through a mapping process. In both stages, the method takes a set of halo properties as input features. By applying the method to a given halo catalog, one can create a subhalo catalog that includes phase space coordinates and values of several other properties for each of its subhalos. It should be noted that in this work we focus on halos/subhalos with $M_{\text{200c}}>10^{11} h^{-1} \text{M}_{\sun}$  with $M_{\text{200c}}$ defined as the mass enclosed within a volume such that the corresponding average dark matter density is 200 times the critical density of the universe. Typically,  $M_{\text{200c}}$  and virial mass values for halos/subhalos are very close to each other. 

This paper is organized as follows. In section \ref{sec:method}, we explain the method in detail. Section \ref{sec:simulation_data} introduces the simulation boxes used in this work and also provides some details about the training, evaluation and test sets. In Section \ref{sec:host_param_evaluation}, we apply the method to two different evaluation sets and evaluate the performances of the method (statistically) when different sets of halo properties are considered as input features for the method. In section \ref{sec:tests}, we  apply the method to three different test sets and create subhalo populations for them and compare a number of statistics measured based on the predicted and actual subhalo populations/sub-populations. The statistics considered here are average subhalo numbers as a function of different halo properties, normalized distributions of subhalos residing in halos of given mass ranges as a function of different subhalo properties and $3$D and projected two point correlation functions for various subhalo samples. In section \ref{sec:discussions}, we present a discussion on improvement and applications of the method and some relevant  issues. Finally, we conclude the paper by summarizing the results in section \ref{sec:conclusion}. 
 
\section{Explaining the method}
\label{sec:method}

Let's assume that we have two sets of halo catalogs, $C_{\text{A}}$ and $C_{B}$, extracted from dark matter simulation boxes ,$S_{\text{A}}$ and $S_{\text{B}}$, where only $C_{\text{A}}$ provides information about subhalos (which can also include sub-subhalos, etc.). In other words, subhalos are absent in $C_{\text{B}}$. Both catalogs provide information about properties of their respective halos (and subhalos in case of $C_{\text{A}}$) which include phase space coordinates and several other properties. We assume that the simulation boxes have the same cosmology, although their initial conditions can be different. In the following two subsections, we introduce a method that first predicts which halos in $C_{\text{B}}$ are hosts, i.e. hosting subhalos above the given mass threshold, and then populates those predicted host halos with subhalos from $C_{\text{A}}$. 

\subsection{The classification part}
\label{subsec: classification}

The first part of the method pertains to solving a straightforward classification problem. In our example, $C_{\text{A}}$ acts as the training set and $C_{\text{B}}$ as the test set. This can be done by either using ML-based classifiers directly or  building a classifier based on ML-based regressors. After testing several ML algorithms for both options, we found that the second option yields more accurate results for the final predictions of the method. It works by first training a regressor on $C_{\text{A}}$ which can predict the number of subhalos residing in each halo in $C_{\text{B}}$ based on a  set of halo properties given as input features. These properties can be virial mass, peak mass, maximum circular velocity, average tidal forces, concentration, etc. We will select a group of such properties in section \ref{sec:host_param_evaluation}. The regressor is trained on ($X^{\text{train}}_{\text{reg}}$, $Y^{\text{train}}_{\text{reg}}$), where $X^{\text{train}}_{\text{reg}}$ and $Y^{\text{train}}_{\text{reg}}$ are feature matrix and target vector of $C_{\text{A}}$ halos, respectively: the $i$th row of $X^{\text{train}}_{\text{reg}}$  contains the values of the input features for the $i$th halo in $C_{\text{A}}$ and the corresponding $Y^{\text{train}}_{\text{reg}}[i]$ value determines the subhalo number for that halo (which is zero for ``empty'' halos and nonzero for host halos). After the training, the regressor makes predictions for the subhalo numbers of $C_{\text{B}}$ halos, creating their predicted target vector ($Y^{\text{pred}}_{\text{reg}}$), based on their feature matrix ($X^{\text{test}}_{\text{reg}}$). Once the predictions are done, the halos in the test set ($C_{\text{B}}$) are classified into empty halos and host halos, using the following criterion: for every halo index $i$ in the test set,
\[
\begin{cases}
	\text{if} ~ Y^{\text{pred}}_{\text{reg}}[i] > 0.5:  &\text{the halo is classified as a host halo},  \\ 
	\text{otherwise:}  &\text{the halo is classified as an ``empty''  halo}. 
\end{cases} \nonumber
\]
There is some flexibility in the choice of the threshold value, depending on the algorithm used for the regression, but $0.5$ is typically a suitable option. 

To find the best regressor, we tested several ML algorithms and compared their speeds and also the accuracies of the final results of the method when each of them was used in the classification part. The best option was a decision tree regressor \citep{breiman1984cart} which, in what follows, we explain a commonly implemented version of it. 

In the training phase, the decision tree regression algorithm starts with the whole training data (feature and target variables) as the ``root node''. It then splits the root node into two child nodes with a splitting criterion that is based on the values of one of the feature variables. The criterion and the corresponding feature variable are chosen such that the weighted sum of the variances of the target values in the resulting child nodes is minimized, with the weights defined based on the sizes of the samples in each node. The algorithm then recursively performs such splittings for the new nodes, with each splitting having its own criterion. This process can continue until either there is only one sample per node or a stopping criterion is met, although the former is barely used as it can lead to over-fitting. The last nodes are usually called ``leaf'' nodes and the resulting data structure resembles a ``tree'', hence the name ``decision tree''. In the prediction phase, every feature vector from a test set is passed through the tree where, by following the splitting criteria, it ultimately lands on one of the leaf nodes and the mean value of the target variables (from the training set) in that node is returned as the corresponding prediction value. 

For this work, we used a stopping criterion that sets the minimum number of samples required to be in a leaf node. We found that setting that number to $3$ results in more accurate final predictions. For the implementation of the regressor, we used the ``DecisionTreeRegressor'' class within the ``sklearn.tree'' module that is part of the scikit-learn library \citep{scikit-learn}. In this class, the hyperparameter that sets the value for our stopping criterion has the name ``min\_samples\_leaf '', which we gave it the value $3$. 

\subsection{The mapping part}
\label{subsec: mapping}

Returning to the catalogs $C_{\text{A}}$ (training set) and $C_{\text{B}}$ (test set), let's assume that we have predicted which halos in $C_{\text{B}}$ are host halos, using the first part of the method explained in the previous subsection. We refer to  host halos in $C_{\text{A}}$ as $C_{\text{A, host}}$ and the predicted hosts in $C_{\text{B}}$  as $C^{\text{pred}}_{\text{B, host}}$. The second part of the method pertains to populating the $C^{\text{pred}}_{\text{B, host}}$ host halos with subhalos  from $C_{\text{A}}$.  For this purpose,  we first use a nearest neighbor search algorithm that for each host halo in $C^{\text{pred}}_{\text{B, host}}$, finds a host halo in $C_{\text{A, host}}$ that is the most similar to that in terms of a selected number of features.  As an example, let's assume that those selected features are just virial mass, $M_{\text{vir}}$, and the dimensionless tidal force exerted on the halo averaged over the past dynamical time, $T_{\text{1dyn}}$. Then, for every predicted host halo in $C^{\text{pred}}_{\text{B, host}}$, the nearest neighbor algorithm finds a host halo in $C_{\text{A, host}}$ such that the distance $\sqrt{ (\Delta \tilde{M}_{\text{vir}})^2 + (\Delta \tilde{T}_{\text{1dyn}})^2}$ between the two halos is minimized, where $\tilde{M}_{\text{vir}}$ and $\tilde{T}_{\text{1dyn}}$ denote the scaled (standardized) values of $M_{\text{vir}}$ and $T_{\text{1dyn}}$. The scaling (standardization) is done to avoid biases in distance measurements when the variables are very different in magnitudes. 

We refer to the predicted host halos (in $C^{\text{pred}}_{\text{B, host}}$) as $\emph{target}$ halos and their corresponding matched hosts in $C_{\text{A, host}}$ as $\emph{donor}$ halos. For the implementation of the nearest neighbor algorithm, we used the ``kneighbors'' method of the ``KNeighborsRegressor'' class from the ``sklearn.neighbors'' module that is provided by the scikit-learn library \citep{scikit-learn}. The corresponding hyperparameter that sets the number of nearest neighbors is ``n\_neighbors'' which we set it to  $1$. The nearest neighbor search is fast thanks to the utilization of the KDTree/BallTree indexing structures by the algorithm.  

For every target halo, once a donor halo is found, the subhalos are ``mapped'' from the donor halo into the target halo using the following assumptions: 1) the relative positions and velocities of subhalos with respect to their host center remain unchanged after mapping. 2) other properties are mapped without change. In general, there is freedom to modify these assumptions or calibrate them. However, in this work we consider these for the method. 

To illustrate how the mapping works, let's consider a pair of target and donor halos, $t$ and $d$, whose centers are located at $\vec{x}_{\text{t}} $ and $\vec{x}_{\text{d}}$ and their velocity vectors are $\vec{v}_{\text{t}} $ and $\vec{v}_{\text{d}}$ respectively. We denote the position and velocity vectors of the $m$-th subhalo (with respect to the global origin of coordinates)  living in the donor host as $\vec{x}^{(m)}_{\text{sub, d}} $ and $\vec{v}^{(m)}_{\text{sub, d}}$ respectively. We also denote the set of properties for the $m$-th subhalo in the donor halo as $\Theta^{(m)}_{\text{sub, d}}$ which can include properties like virial mass, accretion mass, maximum circular velocity at accretion and several others. Assuming that the donor halo has $N_{\text{sub,d}}$ subhalos and considering the two aforementioned assumptions for the mapping process, the method assigns $N_{\text{sub,d}}$ subhalos to the target halo, where the position vector, velocity vector and the set of properties of the $m$-th subhalo ($m = 1, 2, \ldots, N_{\text{sub,d}}$ )  assigned to the target halo are given, respectively, by:
\begin{eqnarray}
	\label{mapping-eqs1}
&&\vec{x}^{(m)}_{\text{sub, t}} =  \vec{x}_{\text{t}} + \left(\vec{x}^{(m)}_{\text{sub, d}} - \vec{x}_{\text{d}} \right), \\
\label{mapping-eqs2}
&&\vec{v}^{(m)}_{\text{sub, t}} =  \vec{v}_{\text{t}} + \left(\vec{v}^{(m)}_{\text{sub, d}} - \vec{v}_{\text{d}} \right), \\
\label{mapping-eqs3}
&& \Theta^{(m)}_{\text{sub, t}} = \Theta^{(m)}_{\text{sub, d}} .
\end{eqnarray}
By repeating this process for every target halo in  $C^{\text{pred}}_{\text{B, host}}$, the method creates a subhalo catalog for the test set which includes phase space coordinates and several other properties for each of the subhalos. 

We end this section by briefly discussing three relevant points. First, in both parts of the method, we require a set of  halo properties to be considered as input features. In general, each part can have its own set of input halo properties but in this work we use the same set for both parts which we will select it in the next section. Second, the performance of any ML algorithm chosen for the classification part should always be evaluated after combining it with the mapping part and generating the subhalo populations. This allows one to choose the algorithm that results in the best overall performance, taking into account the interplay among different contributing factors. Finally, when one works with a training set that is based on a simulation box (of side length $L$) with periodic boundary conditions, some of the host halos in that set that are very close to the faces/edges of the periodic box might have subhalos that are seemingly too far from their centers (separated by $\sim L$) according to their position coordinates. This is just an artefact of the periodic boundary conditions that affects subhalos that ``cross'' a face (or an edge) of a periodic box and appear near the opposite face (or edge). If one of such subhalos is mapped into a predicted host halo in a test box, its position might fall outside of the box. One way to avoid this is to correct the positions of the affected subhalos in the training box before applying the method. The other solution is to first create a subhalo population for the test box using the method and then omit subhalos with anomalous positions for measurements/calculations that require subhalo positions. Since only a tiny fraction of subhalos are affected, the second option is usually safe unless one is interested in measuring/calculating quantities that are highly sensitive to the boundary subhalos.

\section{Simulation boxes and data splits}
\label{sec:simulation_data}

To evaluate and test the method, we use halo catalogs extracted from two N-body simulation boxes,  named Uchuu and mini-Uchuu, which are part of the Uchuu simulations set, DR$1$ \citep{2021MNRAS.506.4210I}. The Uchuu box has a comoving side length of $2\,h^{-1}$ Gpc containing $12800^3$ dark matter particles of mass $3.27\times 10^8 h^{-1} \text{M}_{\sun}$.  The mini-Uchuu box has a comoving side length of $400  \,h^{-1}$ Mpc  containing $2560^3$ particles of the same mass. Both simulations use Planck 2015 cosmological parameters \citep{2016A&A...594A..13P}: $\Omega_m = 0.3089,  \Omega_\Lambda = 0.6911, h  = 0.6774, 
\sigma_\text{8} = 0.8159, \Omega_b = 0.0486$ and $n_s = 0.9667$.  In this work, we only consider halos/subhalos with $M_{\text{200c}}>10^{11} h^{-1} \text{M}_{\sun}$, which means every halo/subhalo is composed of at least $\sim 300$ particles. We also work at redshift zero ($z = 0$), although the method can be used for any other redshifts. 

We therefore refer to the catalogs we use in this work as Uchuu--${\text{m11}z\text{0}}$ and mini-Uchuu--${\text{m11}z\text{0}}$. From the data sets, one can create separate catalogs for halos and subhalos and use the information provided in them, like the ID of the most massive host halo for each subhalo, to make connections between subhalos and their host halos. According to the data, the Uchuu--${\text{m11}z\text{0}}$  has $229336815$ halos which include $16248688$ host halos (the rest being empty of subhalos with masses above the threshold) and a total of $35029728$ subhalos (which also include ``sub-subhalos'', etc. which constitute a small fraction of all subhalos) residing in the host halos. Similarly, the mini-Uchuu--${\text{m11}z\text{0}}$ catalog has $1828932$ halos which include $128024$ host halos and a total of  $275618$ subhalos residing in the hosts. 

In this work we consider two pairs of train-evaluation sets and three pairs of train-test sets that are used in the next two sections for evaluating and testing the method. The evaluation sets and two of the test sets are separated, by random shuffling and with no overlap, from Uchuu--${\text{m11}z\text{0}}$ halos, containing $15\%$,  $15\%$,  $30\%$ and  $30\%$ of the entire halo catalog respectively. For the third test set, we use all halos in the mini-Uchuu--${\text{m11}z\text{0}}$ catalog. We use one training set for both evaluation sets which is the corresponding remaining $70\%$ of the halos, making  $(70\%, 15\% + 15\%)$ train-evaluation sets. For each of the first two test sets, the corresponding remaining $70\%$ of halos is used as the training set, making two $(70\%, 30\% )$ train-test pairs. For the third test set, we use all halos in Uchuu--${\text{m11}z\text{0}}$  as the training set, making a ($100\%$ Uchuu--${\text{m11}z\text{0}}$, $100\%$ mini-Uchuu--${\text{m11}z\text{0}}$) train-test pair.

\section{Selecting the input halo properties}
\label{sec:host_param_evaluation}

In this section, we aim to find which groups of input halo properties (input features) for the method, among the ones we consider here, result in more (statistically) accurate predictions by the method. To this end, we work with the two pairs of train-evaluation sets mentioned in the previous section. We create several subhalo populations for both evaluation sets by the method, using several groups of input halo properties, and compare the statistical accuracies of the predictions as will be discussed later in this section. 

Following the explanations in section \ref{sec:method}, for the $i$th halo in an evaluation set, the method first predicts if it is empty or a host by taking as input a set of its properties. If it is predicted to be empty, the method sets the total number of subhalos it has to zero, i.e. $Y^{\text{pred}} [i] = 0$. If it is predicted to be a host (target) halo, the method uses its mapping algorithm (based on the same set of input properties as the classification part) to predict the total number of subhalos it hosts\footnote{That is the number of the mapped subhalos for that predicted host.}, $Y^{\text{pred}} [i]$, and also the position vector, velocity vector and a set of other properties for each of the predicted subhalos using Eqs \eqref{mapping-eqs1} - \eqref{mapping-eqs3}, which we denote as: $[\vec{x}^{(m)}_{\text{sub}, i}, \vec{v}^{(m)}_{\text{sub}, i}, \Theta^{(m)}_{\text{sub}, i}]$ with $m =   1, 2, ..., Y^{\text{pred}} [i] $. For all of these predictions, the method is trained on the training set for the classification part and also uses the information of the host halos and their subhalos ($[\vec{x}, \vec{v}, \Theta]$) in the training set for the mapping part.

In general, there is no limit on how many subhalo properties can be included in $\Theta_{\text{sub}}$. However, in this work we focus on six subhalo properties that are among preferable candidates for being used as subhalo abundance matching proxies. In all subhalo populations that we create by the method in this work, each subhalo has the following quantities:
\begin{eqnarray}
	\label{sub-predictions-1}
	 &&\vec{x}_{\text{sub}}, \,\, \vec{v}_{\text{sub}}, \\ \label{sub-predictions-2}
	 &&\Theta_{\text{sub}} = \left\{M_{\text{vir}},  V_{\text{max}},    M_{\text{acc}}, V_{0.5}, V_{\text{acc}}, V_{\text{peak}}\right\},
\end{eqnarray}
where $M_{\text{vir}}$ is the virial mass, $V_{\text{max}}$ is the maximum circular velocity, $M_{\text{acc}}$ is the subhalo virial mass at the time of  accretion, $V_{0.5}$ is a special case of $V_{\alpha}$ defines as:
\begin{eqnarray}
	\label{valpha}
	V_{\alpha} = V_{\text{vir}} \left(\frac{V_{\text{max}}}{V_{\text{vir}}} \right)^{\alpha},
 \end{eqnarray}
where $V_{\text{vir}} = (G M_{\text{vir}}/R_{\text{vir}})^{1/2}$ is the virial velocity of the subhalo with $R_{\text{vir}}$ being the virial radius and $\alpha$ being a number which can have any value but is typically chosen to be beween $0$ and $1$, $V_{\text{acc}}$ is the maximum circular velocity of the subhalo at the time of accretion and $V_{\text{peak}}$ is the highest $V_{\max}$ the subhalo has achieved over its entire history.  

In general, including subhalo properties beyond virial mass has the advantage of diversifying the options one can have for connecting subhalos and satellite galaxies via methods like SHAM as one can incorporate various subhalo and satellite galaxy properties. Apart from that, parameters like $M_{\text{acc}}, V_{\text{acc}}, V_{\text{peak}}$ and  $V_{\text{max}}$ are among popular choices for proxies in SHAM models for different reasons. For instance, one reason is the robustness of such properties against tidal stripping effects compared to virial mass (e.g. \cite{2004ApJ...609...35K, 2005ApJ...618..557N, 2006ApJ...647..201C}). We also included $V_{0.5}$ because quantities like $V_{\alpha}$ (defined in Eq.\eqref{valpha}) with different values of $\alpha $ have been used to investigate assembly bias by matching more than one subhalo properties to a galaxy property using abundance matching \citep{2017ApJ...834...37L} \footnote{In that paper, $M_{\text{peak}}$ (the highest virial mass achieved by a halo/subhalo over its entire history) and $V_{\text{peak}}$ are used instead of $M_{\text{vir}}$ and $V_{\text{max}}$ for $V_{\alpha}$ calculations.}. Here, we set $\alpha = 0.5$ as it is one of the most common choices.
\begin{table}
	\centering
	\begin{tabular}{l r}
		Subhalo sample index & Selection criterion \\
		\hline
		$n = 1$  & all subhalos \\
		\hline
		$n = 2$ &  $11 < \log (M_{\text{vir}}/(\text{M}_{\sun}/h)) < 11.5$\\
		\hline
		$n = 3$ &  $11.5 \leq \log (M_{\text{vir}}/(\text{M}_{\sun}/h)) \leq 12$\\
		\hline
		$n = 4$ &  $\log (M_{\text{vir}}/(\text{M}_{\sun}/h)) > 12$ \\
		\hline  
		$n = 5$ &  $V_{\text{max}} < 120$ km/s \\
		\hline
		$n = 6$ &  $120\, \text{km/s}   \leq V_{\text{max}} \leq 180$ km/s \\
		\hline
		$n = 7$  &  $V_{\text{max}} > 180$ km/s \\
		\hline
		$n = 8$ &  $\log (M_{\text{acc}}/(\text{M}_{\sun}/h)) < 11.5$\\
		\hline
		$n = 9$ &  $11.5 \leq \log (M_{\text{acc}}/(\text{M}_{\sun}/h)) \leq 12$\\
		\hline
		$n = 10$ &  $\log (M_{\text{acc}}/(\text{M}_{\sun}/h)) > 12$ \\	
		\hline	
		$n = 11$ &  $V_{0.5} < 100$ km/s \\
		\hline
		$n = 12$ &  $100 \, \text{km/s}  \leq V_{0.5} \leq 150$ km/s \\
		\hline
		$n = 13$  &  $V_{0.5} > 150$ km/s \\
		\hline  
		$n = 14$ &  $V_{\text{acc}} < 120$ km/s \\
		\hline
		$n = 15$ &  $120 \, \text{km/s}  \leq V_{\text{acc}} \leq 180$ km/s \\
		\hline
		$n = 16$  &  $V_{\text{acc}} > 180$ km/s \\
		\hline  
		$n = 17$ &  $V_{\text{peak}} < 150$ km/s \\
		\hline
		$n = 18$ &  $150 \, \text{km/s}  \leq V_{\text{peak}} \leq 220$ km/s \\
		\hline
		$n = 19$  &  $V_{\text{peak}} > 220$ km/s \\
	\end{tabular}
	\caption{Description of subhalo samples for which correlation functions are calculated. For each value of $n$, the corresponding subhalo sample consists of subhalos that satisfy the condition shown on its right side.} 
	\label{tab:sub-selection-inds}
\end{table}
\begin{figure}
	\centering
	\includegraphics[width= \textwidth]{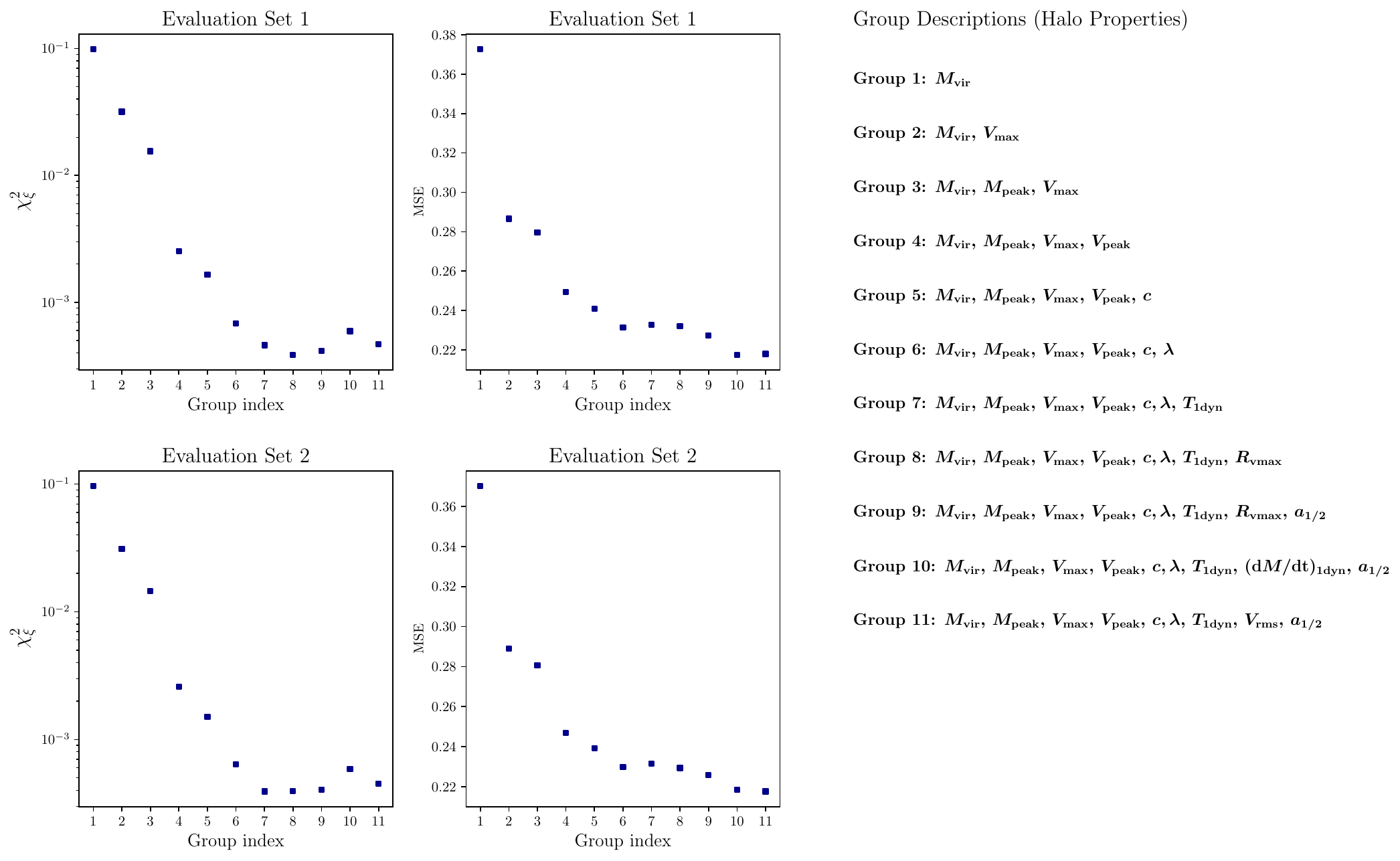}
	\caption{Plots of $\chi^{2}_{\xi}$ (defined in \eqref{chisqrd-xi-def}) and the mean squared error (defined in \eqref{mse-def}) calculated for different subhalo populations created by the method, based on different groups of input halo properties (shown on the right side) for the two evaluation sets.}
	\label{fig:mse-chisquared}
\end{figure}

For each evaluation set, after using the method to generate a \mbox{subhalo} population for a given group of input halo properties, we use two quantities to evaluate the statistical accuracy of the predictions. The first one is the mean squared error between the predicted and actual subhalo numbers of halos, defined as:
\begin{eqnarray}
	\label{mse-def}
	\text{MSE} \equiv \frac{1}{N_{\text{halo}}}\sum_{i = 1}^{N_{\text{halo}}} \left(Y^{\text{pred}}[i] - Y^{\text{actual}}[i] \right)^{2},
\end{eqnarray}
where $N_{\text{halo}}$ is the total number of halos in the evaluation set considered and ``pred'' always refers to ``prediction''. This quantity is sensitive to the accuracy of the predictions for subhalo abundances. For the second quantity, we first select $19$ subhalo samples from each of the predicted and actual subhalo populations which we label them by the index $n$ ($n = 1, 2, ..., 19$). The first sample, $n = 1$, corresponds to the entire subhalo population and $n = 2, ..., 19$ correspond to \mbox{subhalo} samples that are selected based on different ranges of the subhalo properties mentioned in the set \eqref{sub-predictions-2}. For example, $n = 2, 3, 4$ correspond to subhalo samples whose virial masses fall in the ranges $11 < \log (M_{\text{vir}}/(\text{M}_{\sun}/h)) < 11.5$, $11.5 \leq \log (M_{\text{vir}}/(\text{M}_{\sun}/h)) \leq 12$ and $\log (M_{\text{vir}}/(\text{M}_{\sun}/h)) > 12$ respectively. Similarly, for other \mbox{subhalo} properties, we divide their values into three distinct ranges and use them as conditions for selecting the subhalo samples. Therefore, given that there are six properties in the set \eqref{sub-predictions-2}, we will have a total of $18$ subhalo samples which correspond to $n = 2, ..., 19$. We have shown all of the $19 $ indices and their corresponding selection criteria in Table \ref{tab:sub-selection-inds}. We then define the second quantity as:
\begin{eqnarray}
	\label{chisqrd-xi-def}
	\chi^{2}_{\xi} \equiv \frac{1}{285} \sum_{n = 1}^{19} \left [\sum_{j = 1}^{15}\left( \frac{\xi^{ \text{pred}}_{\text{sub}, n}(r_{j}) - \xi^{\text{actual}}_{\text{sub}, n}(r_{j})}{\xi^{\text{actual}}_{\text{sub}, n}(r_{j})}  \right)^2 \right],
\end{eqnarray}
where $\xi^{ \text{pred}}_{\text{sub}, n}(r_{j})$ denotes the subhalo two-point correlation function at the average separation $r_{j}$ (in the bin $j$) for the subhalo sample with index $n$ that is selected from the predicted subhalo population, based on the corresponding selection criterion that can be found in Table \ref{tab:sub-selection-inds}.  Likewise, $\xi^{ \text{actual}}_{\text{sub}, n}(r_{j})$ denotes the similar quantity for the subhalo sample with index $n$ that is selected from the actual subhalo population. In this definition, $j = 1, ..., 15$ correspond to $15$ logarithmic bins from $0.1\, h^{-1}\text{Mpc}$ to $40\, h^{-1}\text{Mpc}$ that determine the subhalo pair separations for two-point function calculations. Defining the second quantity, $\chi^{2}_{\xi}$, in this manner provides us with a way to investigate the accuracy of the clustering of the predicted subhalos in a holistic way, incorporating various subhalo samples/sub-populations and also calculating the correlation functions at different scales spanning both ``one-halo'' and ``two-halo'' regimes. For calculations of  the two point correlation functions, we used the \texttt{Corrfunc} package \citep{2020MNRAS.491.3022S}. 

Figure \ref{fig:mse-chisquared} shows the the values of MSE and $\chi^{2}_{\xi}$ quantities for both evaluation sets, calculated for different outputs of the method based on different groups of halo properties serving as input features. In this figure, $M_{\text{peak}}$ denotes the ``peak mass'' which is the highest mass that a halo achieves throughout its accretion history, $c = R_{\text{vir}}/R_{s}$ is the halo concentration with $R_{s}$ (scale radius) being the radius at which the logarithmic slope of the halo density profile becomes $-2$,  $\lambda$ is the Bullock spin parameter, defined as \citep{2001ApJ...555..240B}:
\begin{eqnarray}
	\label{Bullock-spin-def}
	\lambda = \frac{J}{\sqrt{2} M_{\text{vir}} R_{\text{vir}} V_{\text{vir}}},
\end{eqnarray}
with $J$ denoting the total halo angular momentum, $T_{\text{1dyn}}$ is the dimensionless tidal force exerted on the halo, averaged over the past dynamical time,  Rvmax is the halo radius corresponding to $V_{\text{max}}$, $a_{1/2}$ denotes the ``half-mass scale'' which is the scale factor at which the mass of the most massive progenitor of the halo reaches half of the halo peak mass and $(\text{d} M/\text{d} t)_{\text{1dyn}}$ is the halo accretion rate averaged over the past dynamical time. 

The first group of input halo properties in Figure \ref{fig:mse-chisquared} is just the halo virial mass, $M_{\text{vir}}$. This is the most natural first guess as it has been shown to be strongly correlated with the average subhalo number count, evolved subhalo mass function, etc. (e.g. \cite{2004ApJ...609...35K, 2008MNRAS.386.2135G}). We can clearly see from the decreasing trends of $\chi^{2}_{\xi}$ and MSE in the figure that considering halo properties beyond virial mass as input features makes the method predictions more statistically accurate. The MSE plots are also consistent with the findings of previous studies about the dependence of subhalo abundances on halo properties beyond mass. For instance, it has been shown that at fixed halo mass, subhalo abundances are correlated with halo formation history and concentration (e.g. \cite{2004MNRAS.355..819G, 2005ApJ...624..505Z, 2015ApJ...810...21M}). The plots also show that the method has the best performances when groups $6$ -- $11$ are used as input features. This implies that one can choose any of those groups and obtain comparable results. Here, we select group $8$ as the method's input halo properties for the next (testing) section and denote it by $\Theta^{\text{method}}_{\text{halo}}$. We therefore have:
\begin{eqnarray}
	\label{theta_halo_model}
	\Theta^{\text{method}}_{\text{halo}} = \left\{M_{\text{vir}},  M_{\text{peak}}, V_{\text{max}},  V_{\text{peak}}, c, \lambda, T_{\text{1dyn}}, R_{\text{vmax}} \right\}.
\end{eqnarray}
 
Among the properties included in this group, some of them are strongly correlated with each other. Examples include $M_{\text{vir}}$-$V_{\text{max}}$ and $M_{\text{vir}}$-concentration \citep{1997ApJ...490..493N, 2001MNRAS.321..559B}.  However, this does not necessarily mean that combining these properties will not add any more information. In fact, the plots show otherwise; adding a halo property that is correlated with one or more previous properties can have a complementary or augmenting effect. Some of the properties in our selected group are also correlated with the formation history of the halo, with concentration being one example (e.g. \cite{2001MNRAS.321..559B, 2002ApJ...568...52W}). Furthermore, $T_{\text{1dyn}}$ is an environment-dependent parameter and $R_{\text{vmax}}$ is strongly sensitive to the halo internal structure. It is important to note that the list of halo properties we investigated in Figure \ref{fig:mse-chisquared} is not exhaustive. For instance, one can include other environment-related quantities or halo shape parameters. However, as will be shown in the next section, the method can create subhalo populations with decent statistical accuracy with input halo features like those in $\eqref{theta_halo_model}$ and additional properties that are not considered in this work can serve as improvements to them. We will discuss this more in section \ref{sec:discussions}. 

\section{Testing the method}
\label{sec:tests}
In this section, we test the method using the three different pairs of train-test sets mentioned in section \ref{sec:simulation_data}. Two of the pairs are $70\%$--$30\%$ train--test splits of the Uchuu--${\text{m11}z\text{0}}$ halo catalog which we refer to the corresponding test sets as ``test set $1$'' and ``test set  $2$''. These two test sets have no overlap with each other and their halos span the whole simulation box. The third pair has the entire Uchuu--${\text{m11}z\text{0}}$ halo catalog as the train set and the entire mini-Uchuu--${\text{m11}z\text{0}}$ halo catalog as the test set. We refer to this test set as ``small box''. Following the discussions in sections \ref{sec:method} and \ref{sec:host_param_evaluation}, we use the method to create subhalo populations for the test sets which contain the information about phase space coordinates and other properties of subhalos, as summarized in \eqref{sub-predictions-1} and \eqref{sub-predictions-2}, using $\Theta^{\text{method}}_{\text{halo}}$ (defined in \eqref{theta_halo_model}) as the set of input halo properties for the method. For each pair of the train-test sets, the entire process of training the method and creating the corresponding subhalo population required only a few CPU hours. This can be made even faster by, for instance, using a GPU to implement the method. In the following sub-sections, we calculate a number of statistical quantities based on the predicted and actual subhalo populations and compare them with each other to assess the performance of the method. 

\subsection{Average number of subhalos}
\begin{figure}
	\centering
	\includegraphics[width = \textwidth]{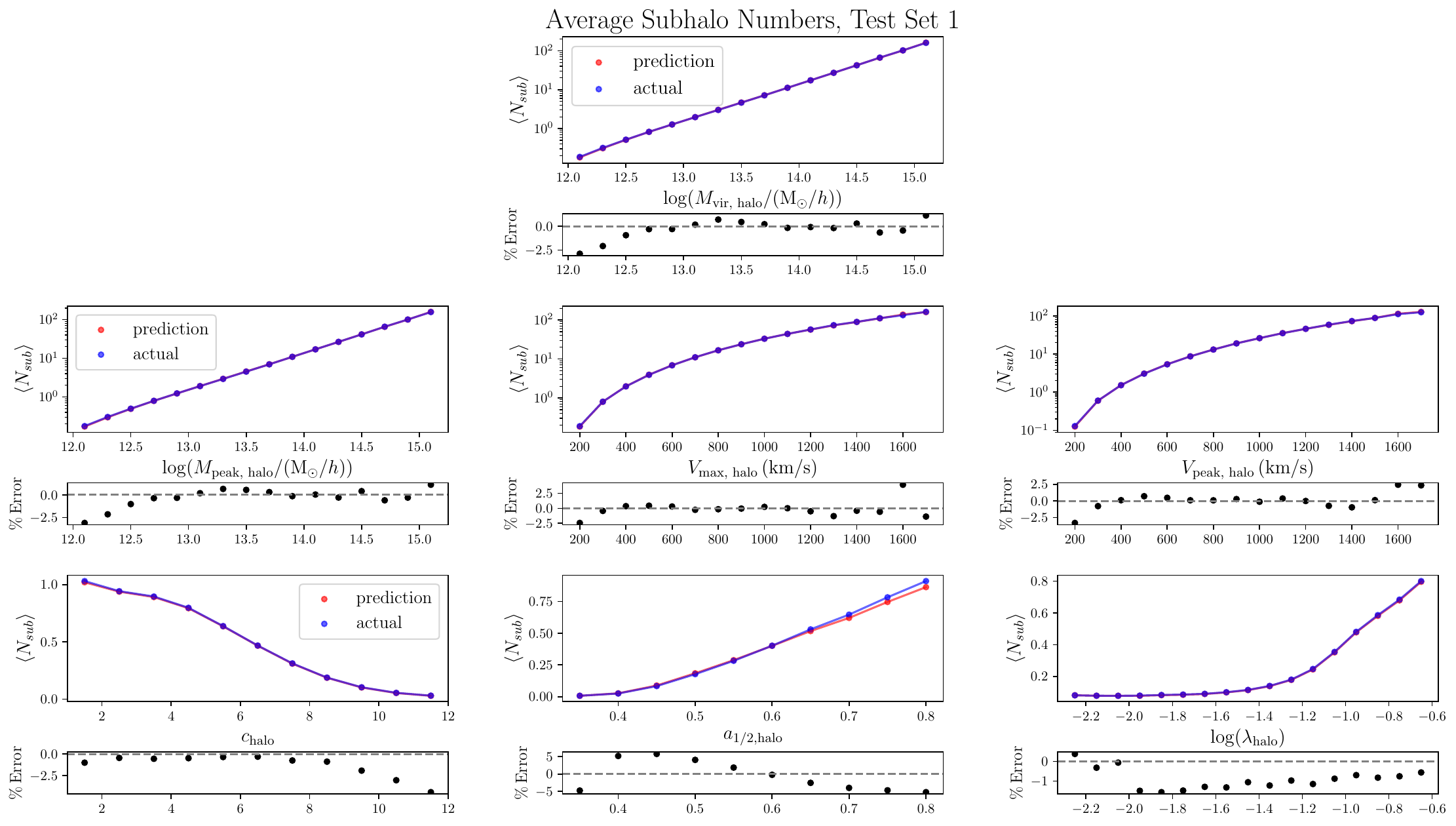}
	\caption{Plots of the average number of subhalos residing in halos of test set $1$ as a function of different binned halo properties, calculated based on the predicted (red) and actual (blue) subhalo populations for test set $1$. The halo properties considered here are virial mass (the topmost plot), peak mass (second row, left), maximum circular velocity (second row, center), peak maximum circular velocity (second row, right), concentration (third row, left), half-mass scale (third row, center) and Bullock spin (third row, right). The percent errors are calculated based on the formula: $\left((\text{predicted} - \text{actual})/\text{actual}\right)\times 100$.}
	\label{fig:Nsub1-avg}
\end{figure}
\begin{figure}
	\centering
	\includegraphics[width = \textwidth]{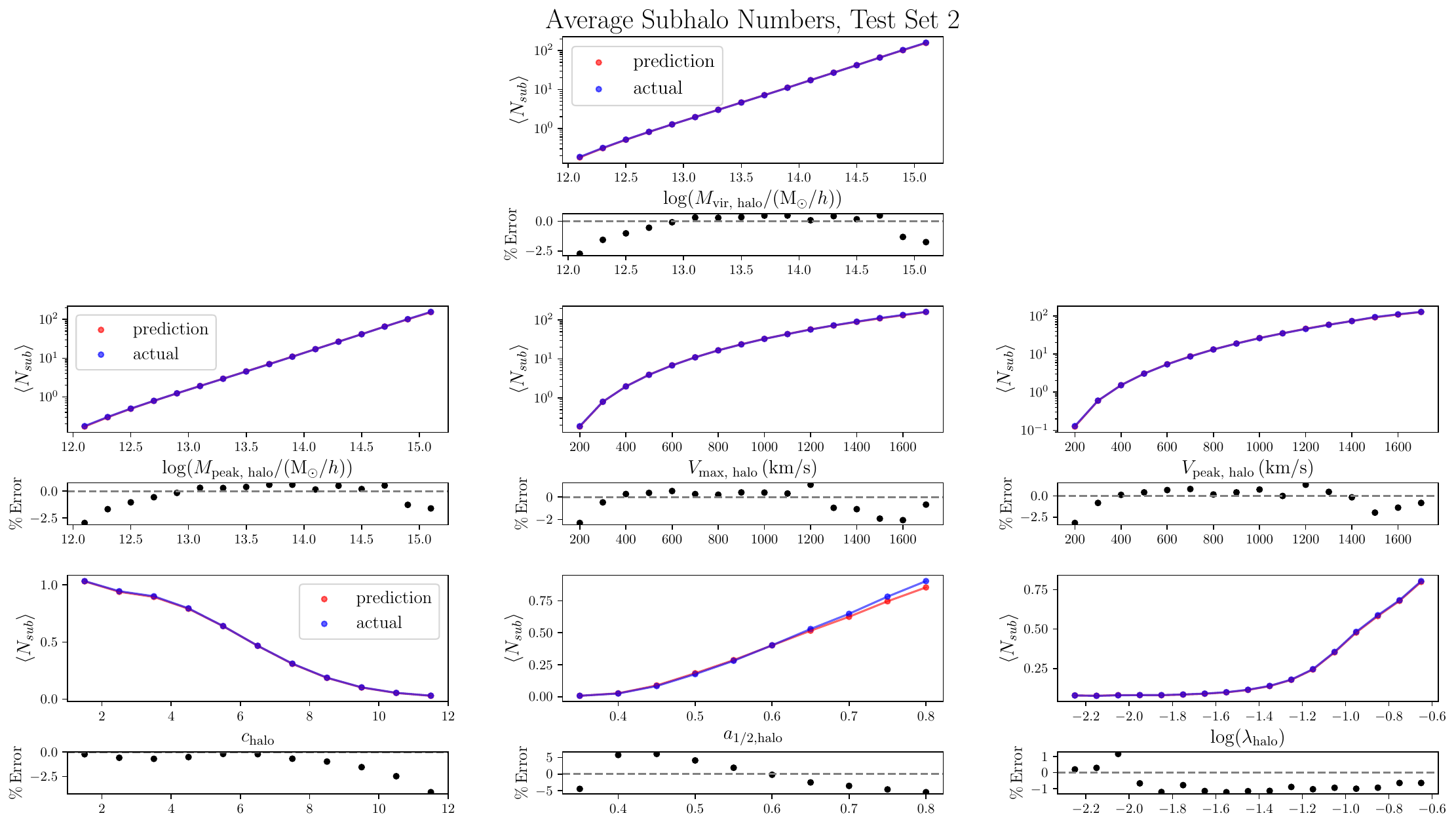}
	\caption{Similar to Figure \ref{fig:Nsub1-avg} but for test set $2$.}
	\label{fig:Nsub2-avg}
\end{figure}
\begin{figure}
	\centering
	\includegraphics[width = \textwidth]{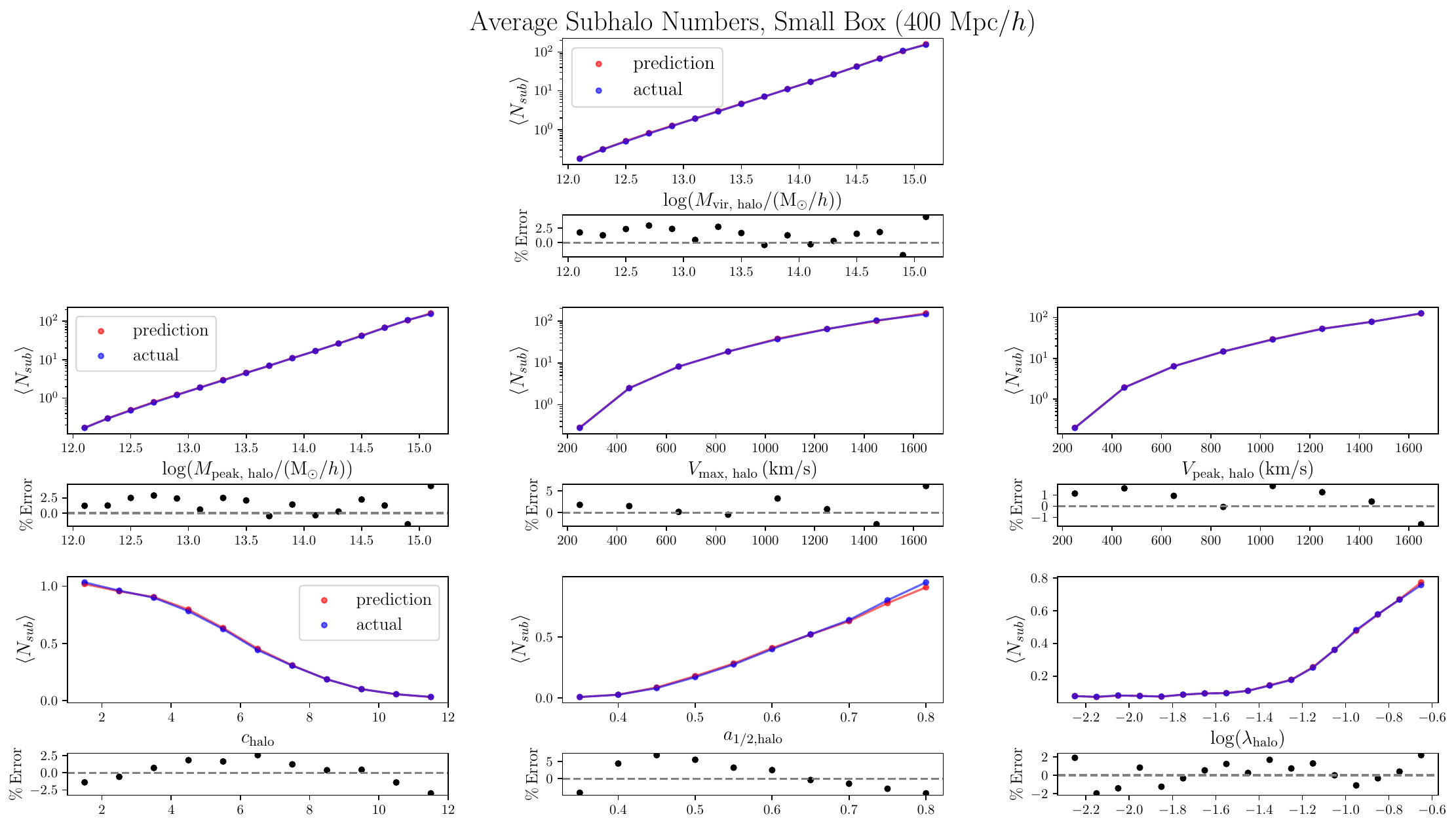}
	\caption{Similar to Figure \ref{fig:Nsub1-avg} but for the small box test set.}
	\label{fig:Nsub-smallbox-avg}
\end{figure}

In this part, we show plots of predicted and actual average subhalo numbers for halos of each test set as a function of different (binned) halo parameters. Figures \ref{fig:Nsub1-avg}-\ref{fig:Nsub-smallbox-avg} show the results, with each figure corresponding to one of the test sets. In each figure, for each bin in a given plot, ``prediction'' and ``actual'' refer, respectively, to the average number of subhalos from the  predicted and actual subhalo populations that reside in halos of that bin. Across all figures, the magnitudes of the percent errors for the vast majority of the predictions are below $5 \%$. It should be noted that since there are far fewer halos in the small box compared to test sets $1$ and $2$, the average subhalo numbers in that set can be more susciptible to Poisson uncertainties compared to their counterparts in the other two test sets. This was especially the case for halos in some of the bins of  $V_{\text{max, halo}}$ and $V_{\text{peak, halo}}$ in the small box test set. To mitigate this, we increased the bin sizes for these two halo parameters which explains why there are fewer points in their corresponding plots in Figure \ref{fig:Nsub-smallbox-avg}. 

\subsection{Normalized distribution of subhalos}
\begin{figure}
	\centering
	\includegraphics[width = \textwidth]{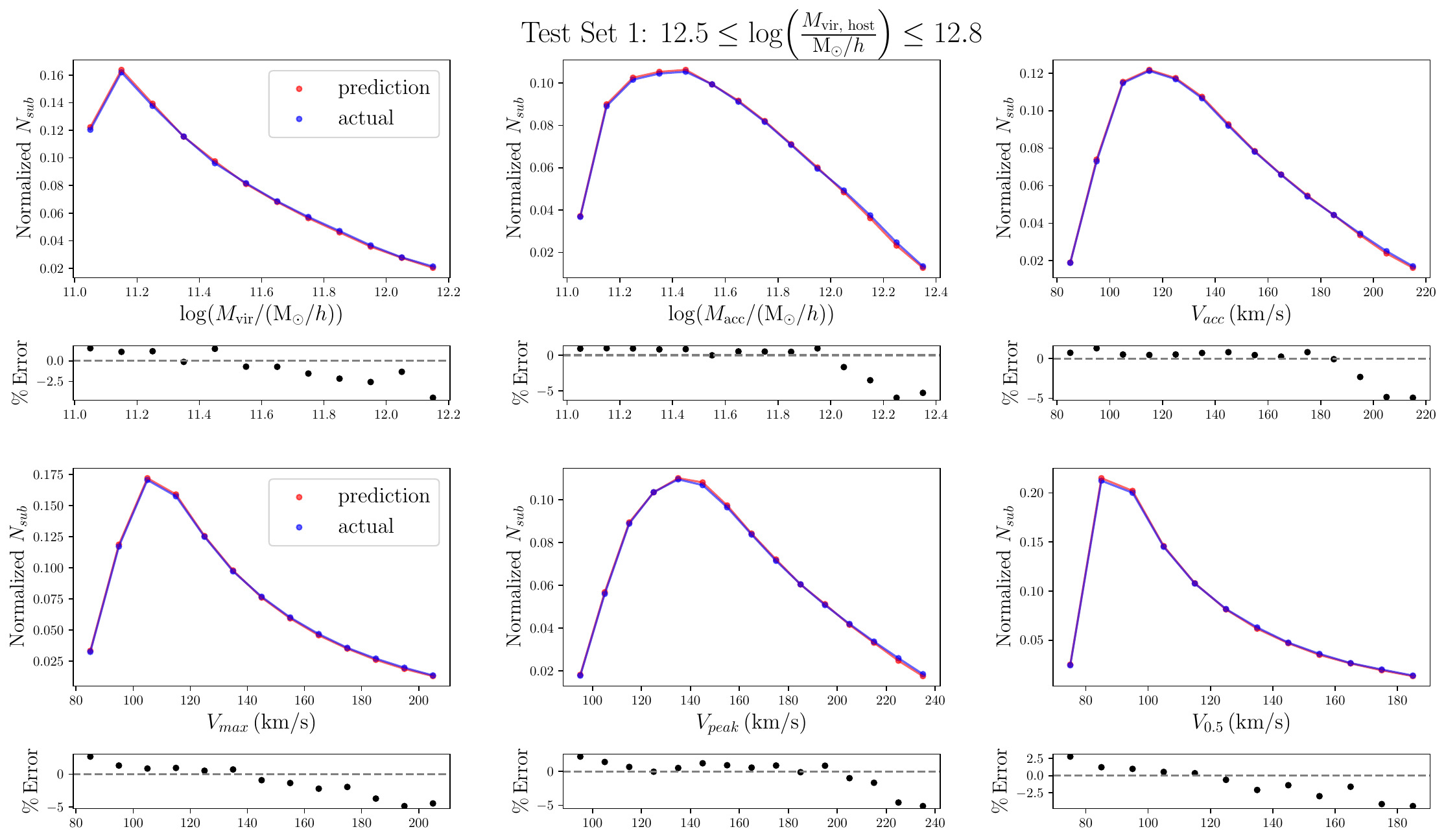}
	\caption{Plots of normalized number of subhalos whose host halos have masses that fall in the range shown in the title, as a function of the (binned) subhalo properties mentioned in \eqref{sub-predictions-2}. The ``prediction'' (red) and ``actual'' (blue) in these plots refer, respectively, to the normalized subhalo numbers that are calculated based on the predicted and actual subhalo populations for test set $1$. The percent errors are calculated based on the formula: $\left((\text{predicted} - \text{actual})/\text{actual}\right)\times 100$.}
	\label{fig:dist1m12}
\end{figure}
\begin{figure}
	\centering
	\includegraphics[width = \textwidth]{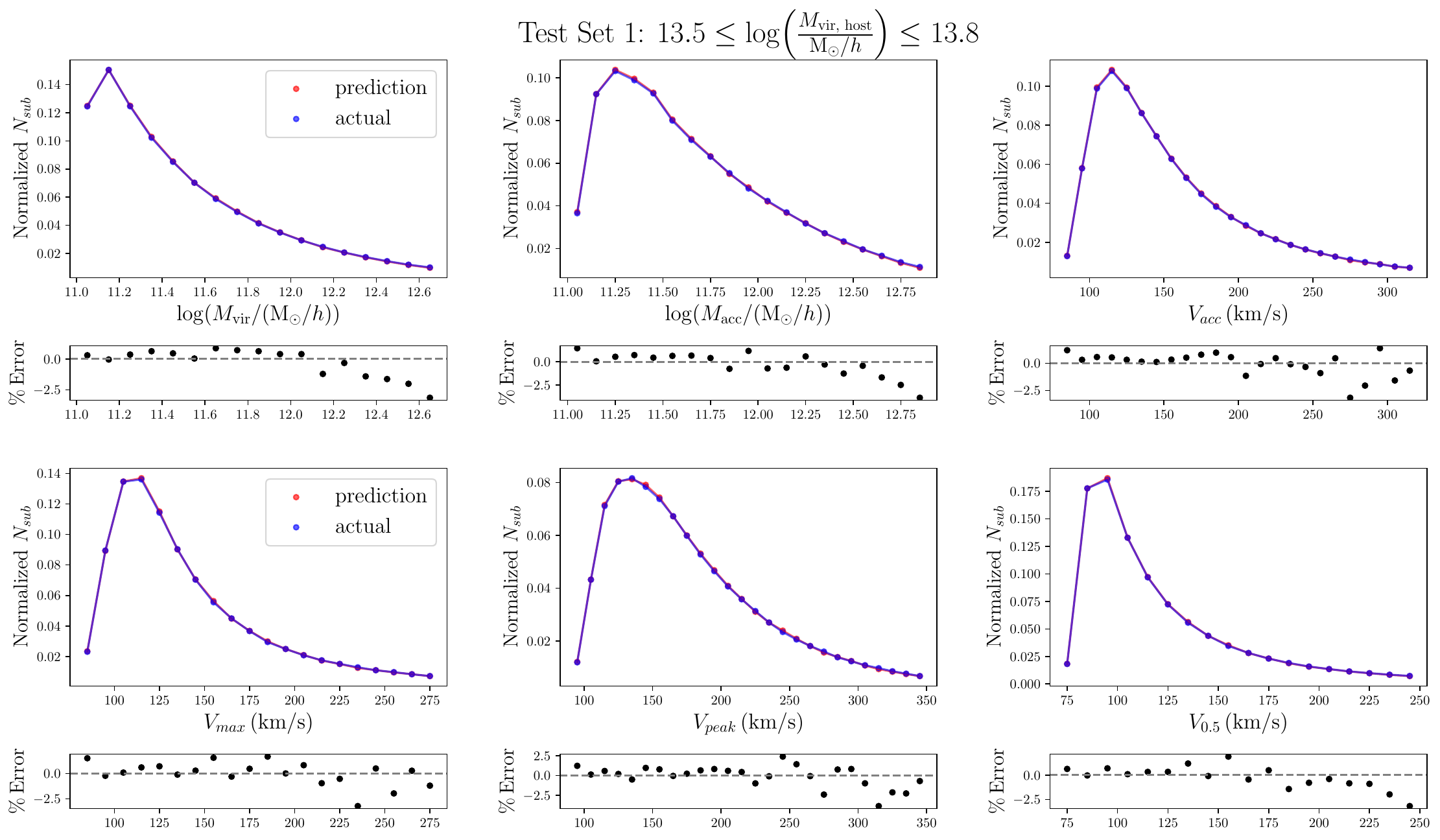}
	\caption{Similar to the previous figure but for a different range of host halo masses, shown in the title.}
	\label{fig:dist1m13}
\end{figure}
\begin{figure}
	\centering
	\includegraphics[width = \textwidth]{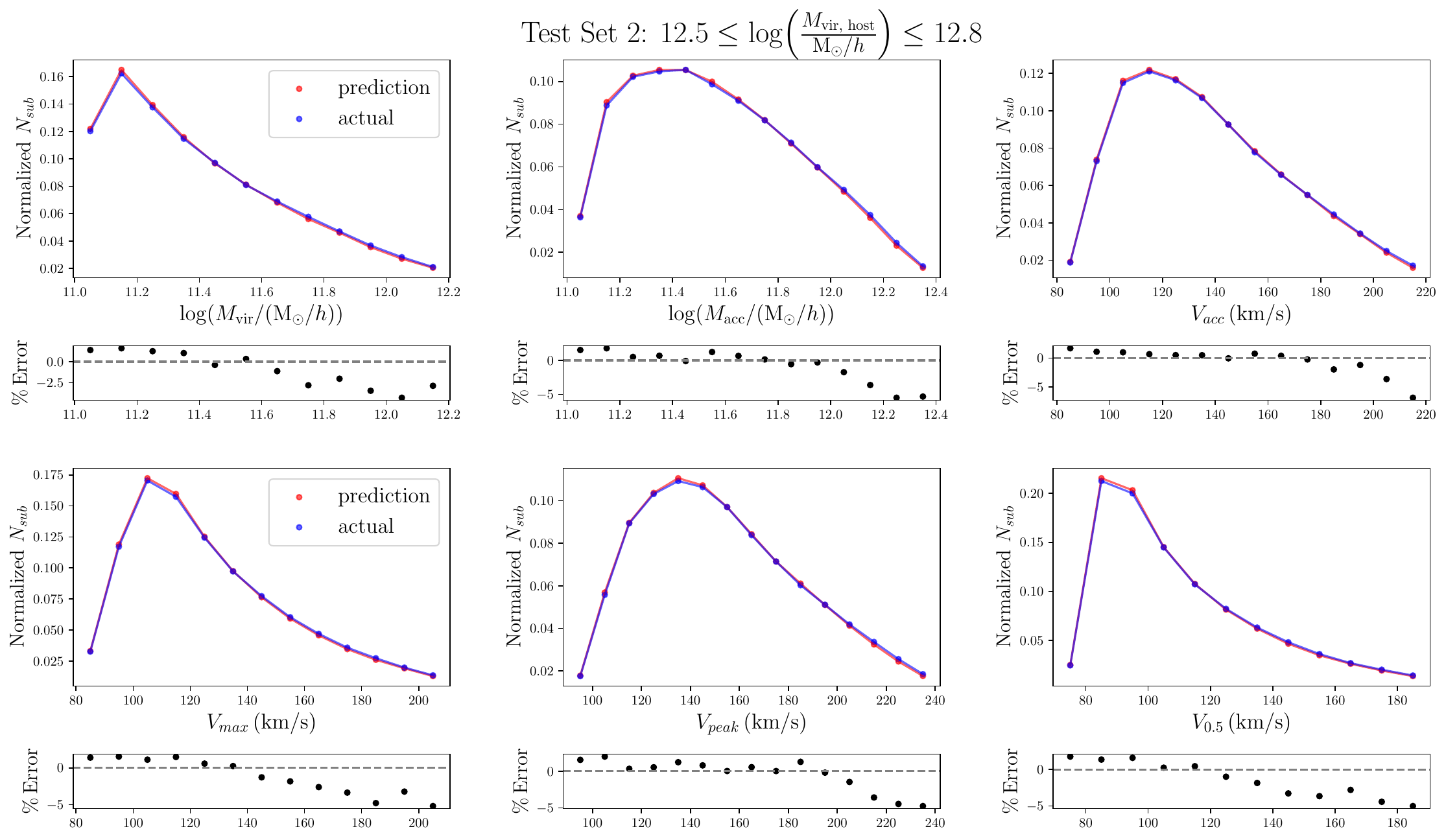}
	\caption{Similar to Figure \ref{fig:dist1m12}  but for test set $2$.}
	\label{fig:dist2m12}
\end{figure}
\begin{figure}
	\centering
	\includegraphics[width = \textwidth]{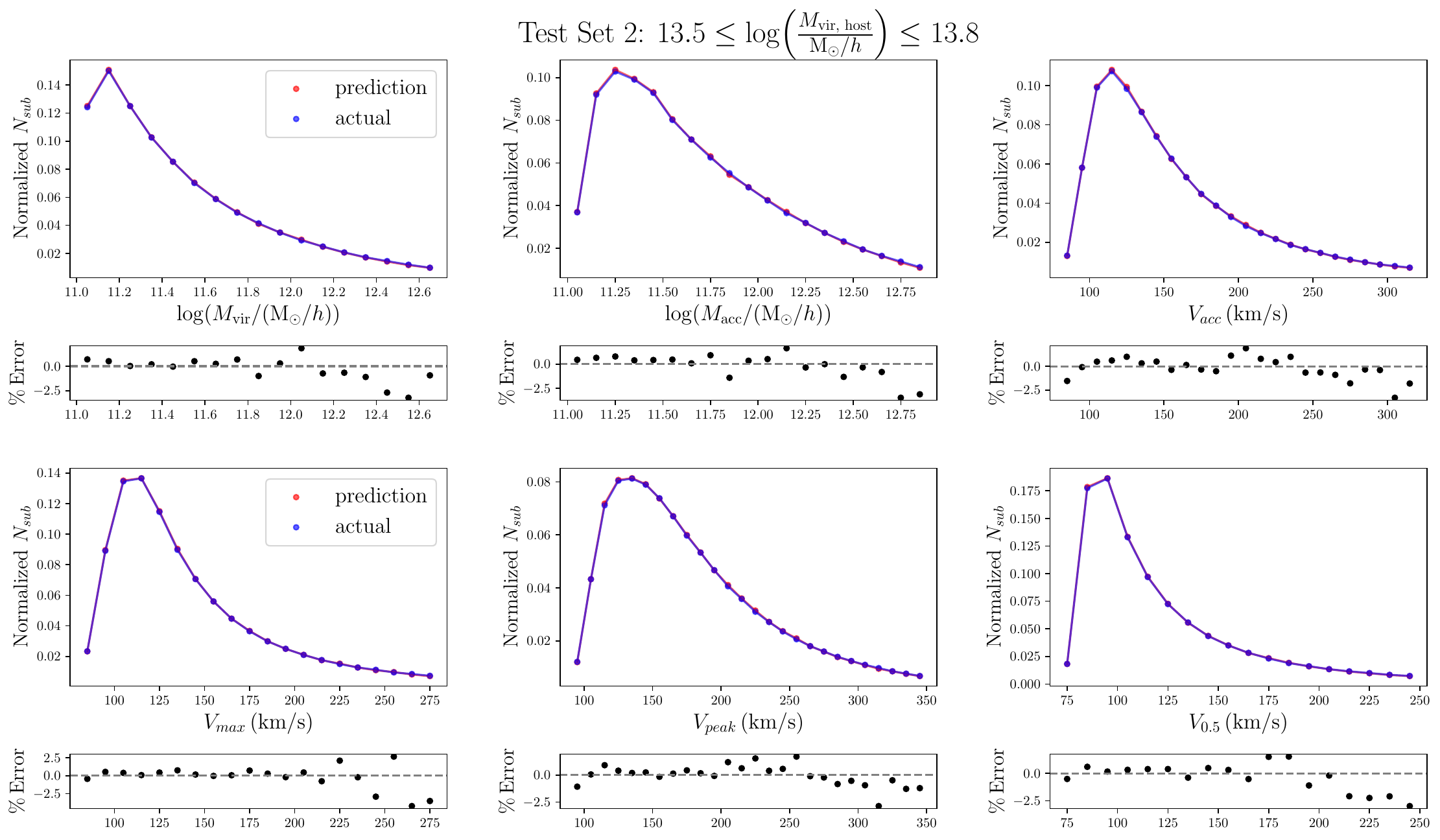}
	\caption{Similar to Figure \ref{fig:dist1m13} but for test set $2$.}
	\label{fig:dist2m13}
\end{figure}

In this part, we show prediction and actual plots of normalized number of subhalos whose host halos have masses that fall in two different ranges, $12.5 \leq \log (M_{\text{vir, host}}/(\text{M}_{\sun}/h)) \leq 12.8$ and  $13.5 \leq \log (M_{\text{vir, host}}/(\text{M}_{\sun}/h)) \leq 13.8$, as a function of different subhalo properties mentioned in \eqref{sub-predictions-2}. For brevity, we refer to these two mass ranges as ``first range'' and ``second range'' respectively. Plots in Figures \ref{fig:dist1m12} and \ref{fig:dist1m13} show the results for test set $1$ and those in Figures \ref{fig:dist2m12} and \ref{fig:dist2m13} show the results for test set $2$. To explain the figures further, let's consider, as an example, the ``prediction'' points (red) in plots of Figure \ref{fig:dist1m12}. To create them, we select all subhalos from the predicted subhalo population for test set $1$ whose host halo masses fall in the first range, forming a sample. Then, for a given bin of a subhalo property, we find the number of subhalos from that sample that fall into that bin and divide it by the total sample size to find the predicted normalized $N_{\text{sub}}$ value for that bin. The``actual'' points (blue) in plots of that figure are also created in a similar way but based on the actual subhalo population of test set $1$. Similar steps are taken for plots of Figures \ref{fig:dist1m13}-\ref{fig:dist2m13}. 

For this part, we only focus on test sets $1$ and $2$ because the relevant subhalo samples in the small box are too small in size that further dividing them into several bins for the distribution plots makes it difficult to make meaningful comparisons between actual and predicted distributions, given the possibility of large Poisson-like fluctuations in the number of subhalos in many of the bins. The sizes of the predicted and actual samples of subhalos whose host halos have masses that fall in the first range are, respectively, $1223742$ and $1229828$ for test set 1 and $1218783$ and $1226155$ for test set 2. Similarly, The sizes of the predicted and actual samples of subhalos whose host halos have masses that fall in the second range are, respectively, $1122398$ and $1119856$ for test set 1 and $1123976$ and $1120711$ for test set 2. In all of these cases the predicted and actual sample sizes have sub-percent differences. 

Considering Figures \ref{fig:dist1m13} and \ref{fig:dist2m13} (corresponding to subhalos with host halo masses in the second range), the magnitudes of the percent errors of all predictions are well below $5 \%$. From Figures \ref{fig:dist1m12} and \ref{fig:dist2m12}, one can see that this is also the case for the vast majority of the predictions, although a few of them in each of these two figures have percent errors whose magnitudes exceed $5\%$. This happens for some of the points that are in the right tails of their respective distributions.  

\subsection{Subhalo clustering statistics}

In this part, we show plots of two-point correlation functions ($3$D) and projected two-point correlation functions ($2$D) for various subhalo samples selected from the predicted and actual subhalo populations for each of the three test sets. The two-point correlation function, $\xi(r)$, of a subhalo sample is a quantity that is proportional to the excess probability, above what is obtained by assuming a random Poisson distribution, of finding two subhalos at a separation distance of $r$. While we only need the real space positions of subhalos to calculate this quantity, in practice, the measured positions of galaxies/subhalos are distorted along the line of sight (LOS) by their peculiar velocities, a phenomenon called ``redshift space distortions'' (RSD) \citep{1972MNRAS.156P...1J, 1987MNRAS.227....1K}. For subhalos in a simulation box at redshift $z_{\text{sim}}$, RSD can be included by mapping the real space position vector, $\vec{x}$, of each subhalo to its ``redshift space'' counterpart, $\vec{s}$, via the relation:

\begin{eqnarray}
	\label{RS-coordinate}
	\vec{s} = \vec{x} + (1 + z_{\text{sim}})\frac{\vec{v}.\mathbf{\hat{x}}}{H(z_{\text{sim}})},  
\end{eqnarray}
where $H(z_{\text{sim}})$ is the Hubble parameter at that redshift and $\vec{v}$ is the velocity of the subhalo.

Considering two redshift space position vectors, $\vec{s}_{1}$ and $\vec{s}_{2}$, the separation vector and the mean LOS vector for the pair are defined as $\vec{s} = \vec{s}_{1} - \vec{s}_{2}$ and $\vec{l} =(\vec{s}_{1} + \vec{s}_{2})/2 $ respectively. Then, the LOS separation, $\pi$, and the projected (transverse) separation, $r_{\text{p}}$, are defined as:
\begin{eqnarray}
	\label{l-and-pi}
	\pi = \frac{\vec{s}.\vec{l}}{|\vec{l}|},~~~~~~  r_{\text{p}} = \sqrt{\vec{s}.\vec{s} - \pi^2}.
\end{eqnarray}
Using these variables and expressing the two-point correlation function in redshift space as $\xi_{s}(r_{p}, \pi)$, the projected correlation function is defined as:
\begin{eqnarray}
	\label{wprp-def}
	w_{p}(r_{p}) = 2 \int_{0}^{\pi_{\text{max}}} \xi_{s}(r_{p}, \pi) d \pi,
\end{eqnarray}
which is a statistic that ``removes'', to some extent, the RSD effects by integrating over the separation along the LOS. The full RSD removal happens when $\pi_{\max} \to \infty$, although in practice much smaller values are chosen and some RSD effects remain. 

\begin{figure}
	\centering
	\includegraphics[width = \textwidth]{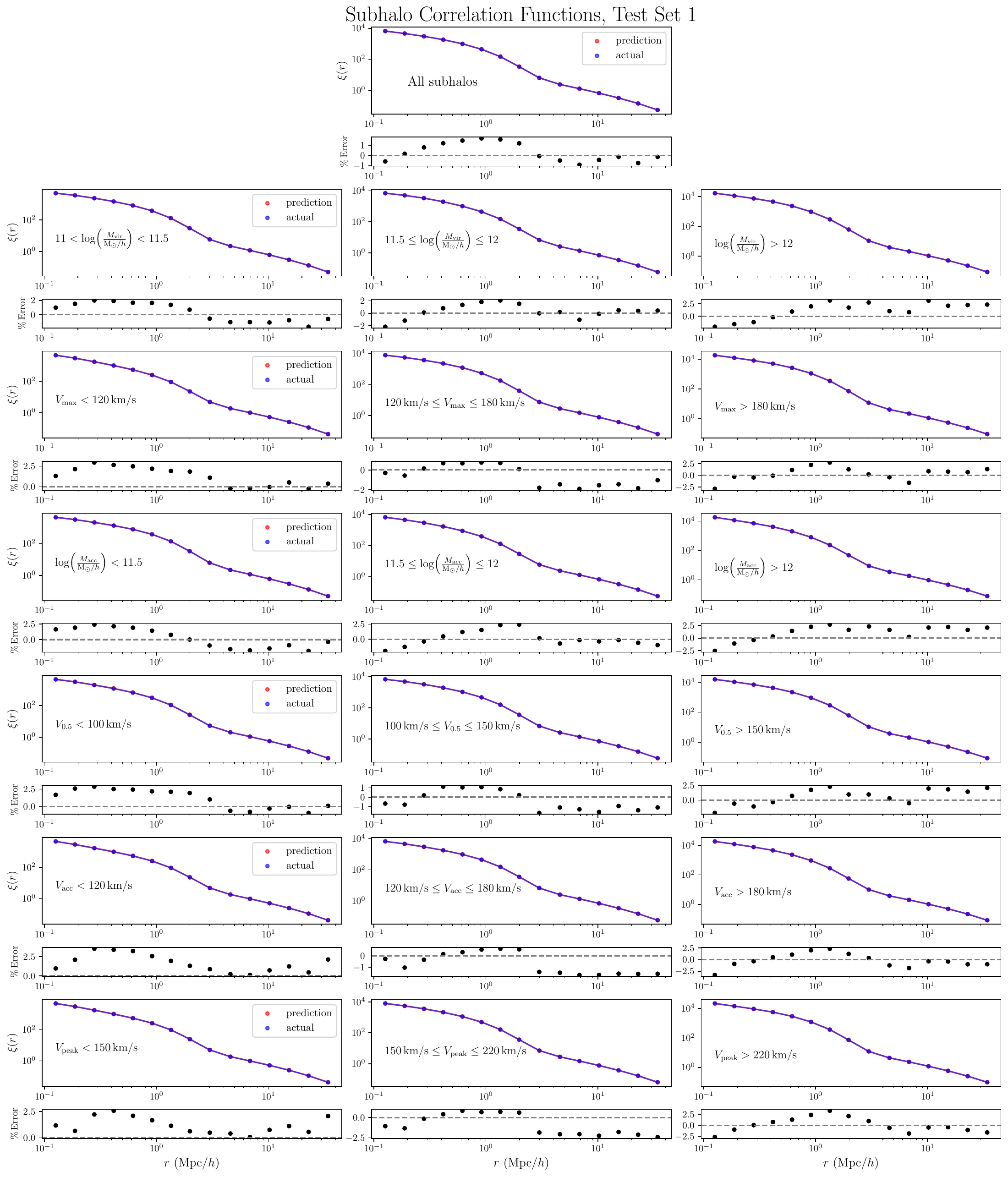}
	\caption{Plots of subhalo two-point correlation functions calculated for various subhalo samples selected from the predicted (red) and actual (blue) subhalo populations for test set $1$. There are $19$ plots corresponding to $19$ predicted and actual subhalo samples whose selection criteria are shown on the plots and can also be found in Table \ref{tab:sub-selection-inds}. The percent errors are calculated based on the formula: $\left((\text{predicted} - \text{actual})/\text{actual}\right)\times 100$.  }
	\label{fig:xi_set1}
\end{figure}
\begin{figure}
	\centering
	\includegraphics[width = \textwidth]{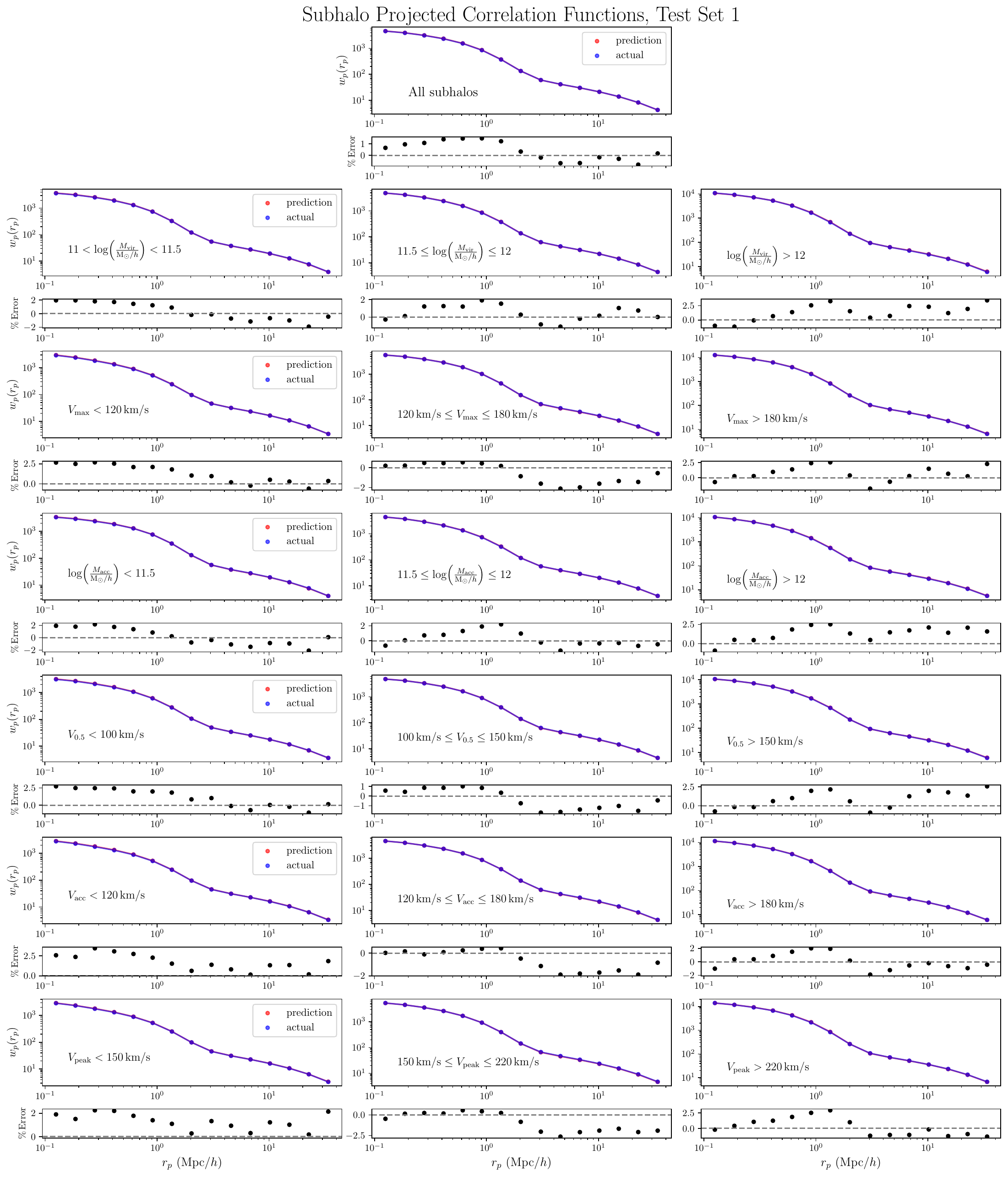}
	\caption{Plots of subhalo \textbf{projected} two-point correlation functions calculated for various subhalo samples selected from the predicted (red) and actual (blue) subhalo populations for test set $1$. There are $19$ plots corresponding to $19$ predicted and actual subhalo samples whose selection criteria are shown on the plots and can also be found in Table \ref{tab:sub-selection-inds}.   }
	\label{fig:wp_set1}
\end{figure}
\begin{figure}
	\centering
	\includegraphics[width = \textwidth]{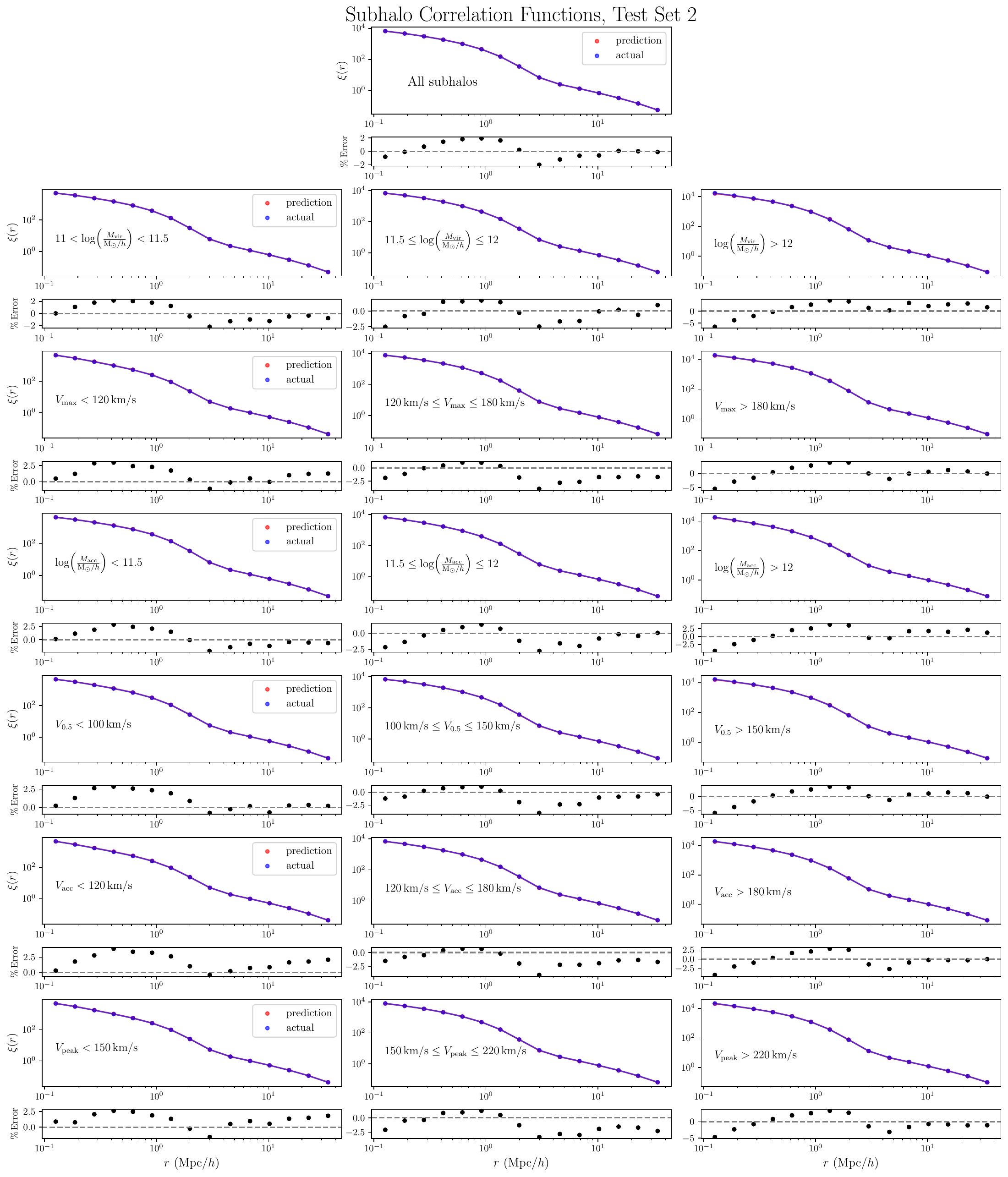}
	\caption{Similar to Figure \ref{fig:xi_set1} but for test set $2$.}
	\label{fig:xi_set2}
\end{figure}
\begin{figure}
	\centering
	\includegraphics[width = \textwidth]{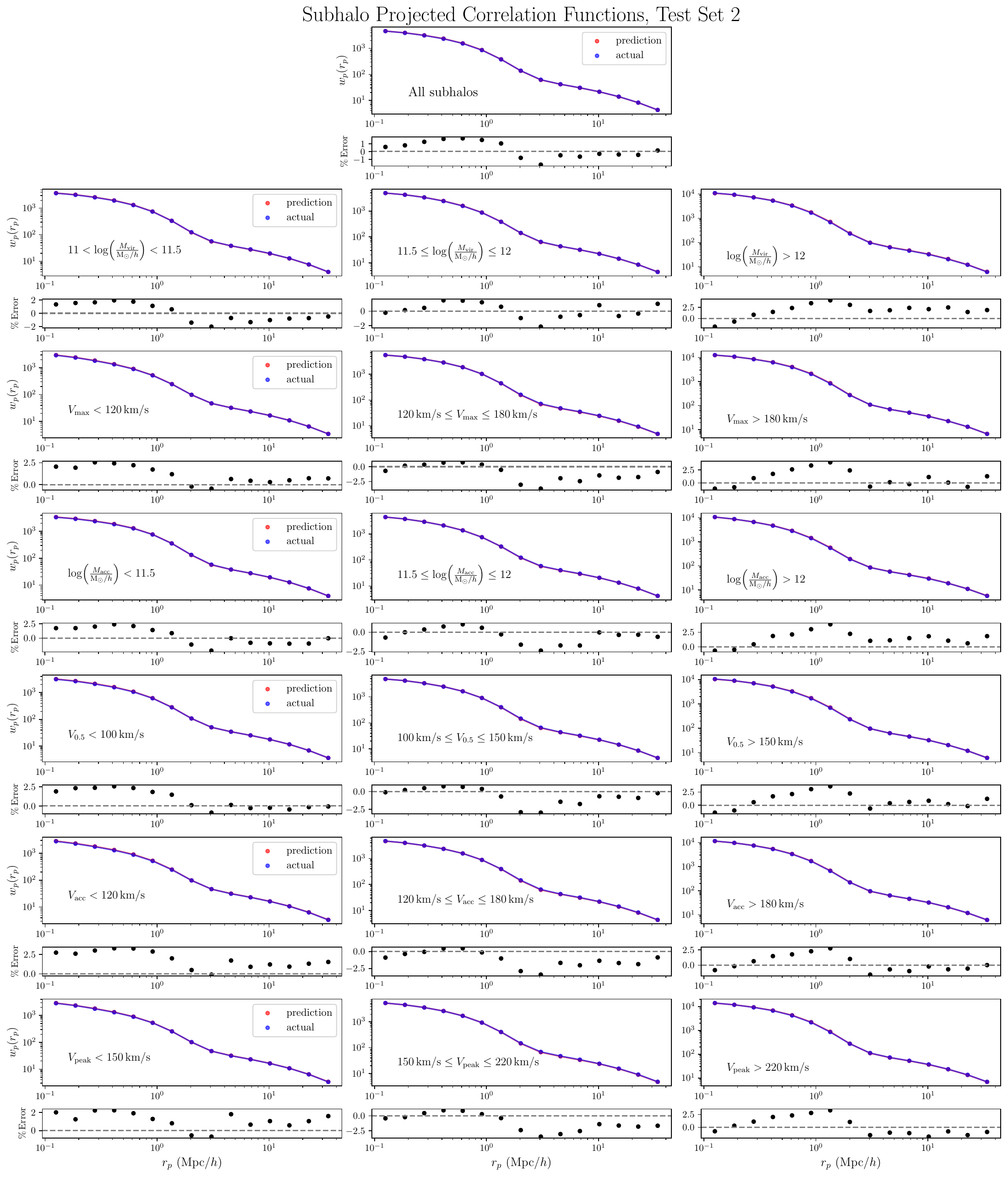}
	\caption{Similar to Figure \ref{fig:wp_set1} but for test set $2$.}
	\label{fig:wp_set2}
\end{figure}
\begin{figure}
	\centering
	\includegraphics[width = \textwidth]{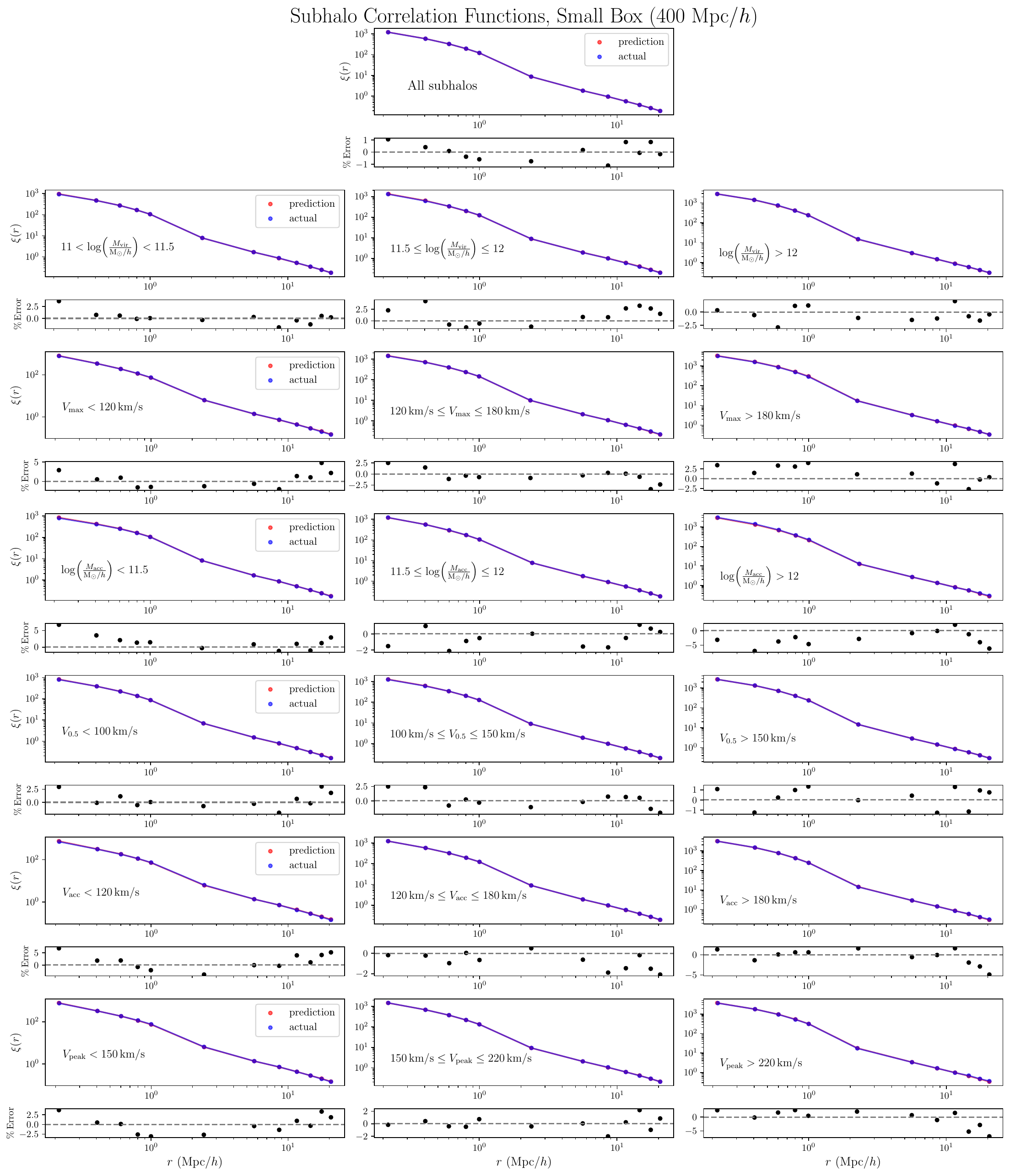}
	\caption{Similar to Figure \ref{fig:xi_set1} but for the small box test set.}
	\label{fig:xi_smallbox}
\end{figure}
\begin{figure}
	\centering
	\includegraphics[width = \textwidth]{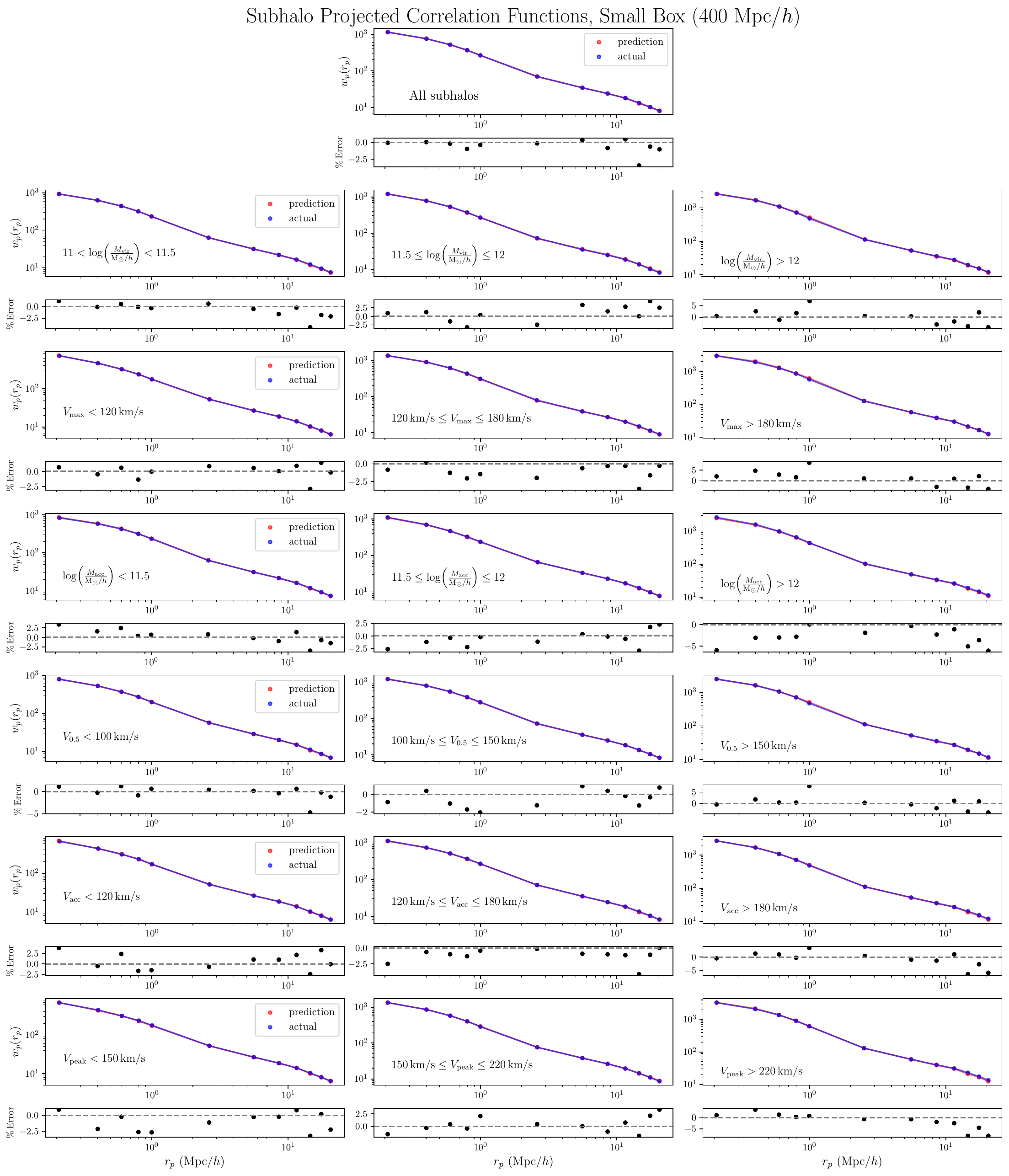}
	\caption{Similar to Figure \ref{fig:wp_set1} but for the small box test set.}
	\label{fig:wp_smallbox}
\end{figure}

Figures \ref{fig:xi_set1}-\ref{fig:wp_smallbox} show the results for values of $\xi(r)$ and $w_{p}(r_{p})$ calculated for various subhalo samples that are selected from the predicted and actual subhalo populations for the three test sets. Figures \ref{fig:xi_set1}, \ref{fig:xi_set2} and \ref{fig:xi_smallbox} show the results for $\xi(r)$ values for test set $1$, test set $2$ and the small box respectively and Figures \ref{fig:wp_set1}, \ref{fig:wp_set2} and \ref{fig:wp_smallbox} show the results for $w_{p}(r_{p})$ values for test set $1$, test set $2$ and the small box respectively. In each figure, the plots show the results for subhalo samples whose selection criteria were described in Table \ref{tab:sub-selection-inds} (and are also shown on the plots). In other words, in each figure, there are $19$ plots that correspond to $n = 1, ..., 19$ subhalo samples in that table; The topmost plot corresponds to $n =1$ (all subhalos), the three plots in the row below that correspond to $n = 2, 3, 4$ (subhalo samples selected based on virial mass) and, following a similar pattern, plots in the next rows correspond to $n= 5, 6, ..., 19$.  

To make the predicted or actual $\xi(r)$ plots, we used the real space position coordinates of subhalos, $\vec{x} = (x, y, z)$, from the corresponding predicted or actual samples. For making the $w_{p}(r_{p})$ plots, one first needs to map the real space coordinates to their redshift space counterparts via Eq. \eqref{RS-coordinate}.  To do this, we assumed that LOS direction is aligned with the $z$ direction which is sufficient for our comparisons in this work. In other words, we performed the following mappings to obtain the redshift space coordinates of subhalos at redshift zero ($z_{\text{sim}} = 0$ in equation \eqref{RS-coordinate}): $s_{x} = x$, $s_{y} = y$, $s_{z} = z + v_{z}/H_{0}$, where $v_{z}$ is the $z$ component of the subhalo velocity vector. Given these coordinates, the projected correlation function obtained from Eq. \eqref{wprp-def} is in fact a projection onto the $x-y$ plane. Having the subhalo coordinates in real and redshift spaces, $\vec{x}$ and $\vec{s}$, sets the stage for calculating $\xi(r)$ and $w_{p}(r_{p})$ for various predicted and actual subhalo samples as discussed before. We used the \texttt{Corrfunc} package \citep{2020MNRAS.491.3022S} for all $\xi(r)$  and $w_{p}(r_{p})$ calculations in this work. 

For test sets $1$ and $2$, we calculated the subhalo two-point functions for $15$ logarithmic bins of the pair separations, from $0.1\, h^{-1}\text{Mpc}$ to $40\, h^{-1}\text{Mpc}$. For the small box test set, we used a different binning for the pair separations, that is $5$ linear bins from $0.1\, h^{-1}\text{Mpc}$ to $1.1\, h^{-1}\text{Mpc}$ with bin size of $0.2\, h^{-1}\text{Mpc}$ combined with $7$ linear bins from $1.1\, h^{-1}\text{Mpc}$ to $22 \, h^{-1}\text{Mpc}$ with bin size of $3 \, h^{-1}\text{Mpc}$. The reason we chose a different binning for the small box was to minimize the Poisson errors caused by small numbers of subhalo pairs in some of the bins. In fact, for a given bin of pair separations, the Poisson error of the corresponding $\xi(r)$ value (with $r$ being the average separation of the pairs in that bin) scales approximately as $1/\sqrt{N_{\text{pair}}(r)}$ where $N_{\text{pair}}(r)$ is the number of pairs in that bin. Therefore, it is preferable to choose separation bins for which $N_{\text{pair}}(r)$ is large enough to make Poisson errors negligible. Altogether, with our chosen sets of bins, the Poisson errors across all measurements in the three test sets become too small to alter the conclusions about the method's performance. Also for each test set, we chose a maximum pair separation that is well below the corresponding box size: $40\, h^{-1}\text{Mpc}$ for test sets $1$ and $2$ whose simulation box has a side length of $2000\, h^{-1}\text{Mpc}$ and $22 \, h^{-1}\text{Mpc}$ for the small box which has a side length of $400\, h^{-1}\text{Mpc}$. For calculations of the projected correlation functions, $w_{p}(r_{p})$, our $r_{p}$ bins are similar to the $r$ bins explained above.  We also set $\pi_{\text{max}} = 40\, h^{-1}\text{Mpc} $ for test sets $1$ and $2$ and $\pi_{\text{max}} = 20\, h^{-1}\text{Mpc} $ for the small box test set.

Considering all plots in Figures \ref{fig:xi_set1}-\ref{fig:wp_smallbox}, the overwhelming majority of the predictions have percent errors whose magnitudes are well below $5\%$. For test set $1$ (Figures \ref{fig:xi_set1} and \ref{fig:wp_set1}), the magnitudes of percent errors for all predictions are below $5\%$. For test set $2$, only $\sim 1\%$ of all $\xi(r)$ predictions (Figure \ref{fig:xi_set2}) and none of the $w_p(r_p)$ predictions (Figure \ref{fig:wp_set2}) have percent errors whose magnitudes exceed $5 \%$. Finally, for the small box test set, $\sim 3 \%$ of all $\xi(r)$ predictions (Figure \ref{fig:xi_smallbox}) and $\sim 4\%$ of all $w_p(r_p)$ predictions (Figure \ref{fig:wp_smallbox}) have percent errors whose magnitudes exceed $5 \%$. One possible reason that there are more such predictions for the small box test set is that there are far fewer subhalos in that set compared to the other test sets. Although, as we discussed before, we do not expect that Poisson uncertainties would change these results significantly, the method, in general, has a better performance (in a statistical sense) for test sets that in reality have more subhalos. Considering the observations from all the figures in this section, we can conclude that the method has an overall satisfactory performance. 

\subsection{A point about the input halo properties for the method}
As it was mentioned before, among the groups of input halo properties for the method that we considered in section \ref{sec:host_param_evaluation}, one can select any of the groups $6-11$ (shown in Figure \ref{fig:mse-chisquared}) without significantly changing the performance of the method. Although we chose group $8$, if we had selected a different group, the subhalo populations created by the method might have had slightly less accurate clustering but, for instance, their abundances might have been slightly more accurate. Since the method is fast in creating subhalo populations, one can test different groups of input halo properties and use the one that results in the most accurate predictions for the specific statistics that are desired to be analyzed. This gives one a high degree of flexibility in using the method for generating subhalo populations for statistical analyses.

\section{Discussion on improvement and applications of the method}
\label{sec:discussions}

Before discussing the method's applications, it should be reminded that the sets of input halo properties we considered in section \ref{sec:host_param_evaluation} for comparing the performances of the method (Figure \ref{fig:mse-chisquared}) is not exhaustive and the selected set in \eqref{theta_halo_model} can be expanded. Possible additional parameters can be those that are related to the halo environment and/or shape. For instance, it has been shown that the spatial distribution of subhalos within their host halo is correlated with the host halo shape (e.g. \cite{2005ApJ...629..219Z, 2005MNRAS.364..424W, 2014MNRAS.442.1197H, 2025MNRAS.538..963M}) and also the abundance and other properties of subhalos are affected by their host halo environment (e.g. \cite{2011MNRAS.413.1973W, 2025A&A...700A..65H}). Some of the halo properties we considered as input features do correlate with halo environment and/or shape. Examples are halo mass and concentration (e.g. \cite{2001MNRAS.321..559B, 2005ApJ...627..647B, 2006MNRAS.367.1781A, 2007MNRAS.378...55M, 2017MNRAS.466.3834L, 2021PhRvD.103f3517H}) and also the (dimensionless) tidal force exerted on halo, averaged over the past dynamical time ($T_{\text{1dyn}}$). However, there can be other quantities that are more directly correlated. Environmental halo properties are especially interesting because some of them can also impact the shapes of the halos (e.g. \cite{2005ApJ...627..647B, 2007MNRAS.375..489H, 2011MNRAS.413.1973W, 2017MNRAS.466.3834L}). Examples of environmental halo properties include spherical density contrasts calculated for different radii around a halo, quantities defined based on the tidal tensor calculated for the local halo environment (that can also be used to quantify environmental anisotropies) and parameters that determine the type of cosmic web structure in which the halo resides (filament, void, etc.). The inclusion of such properties may improve the performance of the method and make the results even more accurate. This is an interesting direction to be explored elsewhere. 

One of the applications of  the method is populating dark matter halos in low resolution simulation boxes with subhalos from a higher resolution simulation box with the same cosmology but arbitrary initial conditions. This can significantly expedite statistical analyses that require a large number of simulation boxes such as estimations of covariance matrices for various quantities that are related to subhalos in a direct or indirect way (e.g. different clustering statistics of subhalos or satellite galaxies whose mock catalogs are created via methods like SHAM). Since the method takes as input a number of halo properties, one issue that might arise is that some halo properties that are measured in a low resolution box might be biased with respect to similar properties measured in the high resolution box due to differences in simulation parameters like force softening scale (e.g. \cite{2021MNRAS.500.3309M}). This is especially the case for lower mass halos in a low resolution simulation box. Although this may not affect the performance of the method in many cases, if need be, one can ``calibrate' the halo parameters in the low resolution boxes in order to reduce/remove such biases. In the next paragraph, we elaborate on this further. 

Suppose that we have a high resolution simulation box, $H$, and we aim to use the method to populate $n$ simulation boxes with much lower resolutions, $L_{1}, L_{2}, ..., L_{n}$, with subhalos from $H$. Each $L_{j}$ simulation has initial conditions that are different from those of other $L_{i \neq j}$ boxes and $H$, although the cosmologies are the same. In order to check if halo properties in the low resolution boxes need to be calibrated, one can run a special low resolution simulation, $L_{\text{cal}}$,  which is similar to $H$ in every aspect except its resolution, which should be similar to that of the $L_{j}$ boxes, and make comparisons between halo populations in $L_{\text{cal}}$ and $H$. For instance, since $L_{\text{cal}}$ is just a low resolution version of $H$, it is expected that rank ordering halos in $L_{\text{cal}}$ and $H$ according to different halo properties yields consistent results. By comparing such rank orders, one can investigate how different the properties of halo populations are in $L_{\text{cal}}$ and $H$ and then, if necessary, introduce halo mass thresholds in order to exclude low mass halos in $H$ that may not be resolved in $L_{\text{cal}}$ and correction factors/functions in order to scale or correct the values of halo properties in $L_{\text{cal}}$ that are affected by a resolution-induced bias. Such possible thresholds and corrections can then be applied to halos in $L_{j}$ boxes and then the method can be used as usual to create subhalo populations. In general, there can be other calibration strategies and here we aimed to mention just one example.

The other application of the method is identifying halo properties that are (more strongly) correlated with subhalo abundance/clustering in a given dark mater simulation box. The idea is to do an analysis similar to what was done in section \ref{sec:host_param_evaluation} (possibly with a larger number of evaluation sets) to find groups of input halo properties that cause the method to generate the most statistically accurate \mbox{subhalo} populations for the given simulation. Such an analysis can in general involve other statistical tests and groups of input halo properties not considered here, although the basic idea is the same. For a given set of statistical quantities that quantify statistical accuracies of the subhalo populations created by the method from different aspects (such as MSE and $\chi^{2}_{\xi}$ we considered in section \ref{sec:host_param_evaluation} and possibly other ones), if
all of such quantities indicate a better performance of the method for a group $A$ of input halo properties compared to a group $B$ of input halo properties by a sufficiently large margin and in all evaluation sets, then one can conclude that the (combination of) halo properties in $A$ are more strongly correlated with subhalo abundance/clustering compared to the (combination of) halo properties in $B$. Ultimately, through galaxy-halo connection models like SHAM, such information can provide valuable insights into the physics of galaxy assembly bias. 

\section{Summary}
\label{sec:conclusion}

In this work we presented a machine learning-based method that first predicts which halos in a given dark matter N-body simulation box host substructure above a given mass threshold and then populates the predicted hosts with subhalos from a high resolution simulation box which has the same cosmology but can have different initial conditions. The method itself was explained in detail in section \ref{sec:method}. In section \ref{sec:simulation_data}, we introduced the simulation boxes used in this work for evaluating and testing the method and explained how the data splittings have been done. In section \ref{sec:host_param_evaluation}, we considered several groups of halo properties as input features for the method and, in each case, we evaluated the statistical accuracies of the predictions  for two different evaluation sets (using two statistical quantities for each evaluation set). The results were shown in Figure \ref{fig:mse-chisquared}. It is evident from that figure that adding halo properties beyond virial mass to the input features results in more accurate predictions by the method. Also in that section, the halo properties in one of the groups, listed in \eqref{theta_halo_model}, were selected as the method's input features for the testing section. 

In section \ref{sec:tests}, using the method, subhalo populations were created for three different test sets, namely test set 1, test set 2 and the small box test set. Then a number of statistical quantities were measured/calculated based on the predicted and actual subhalo populations (and sub-populations/samples from them) and were compared with each other in order to asses the performance of the method. Figures \ref{fig:Nsub1-avg}-\ref{fig:Nsub-smallbox-avg} show plots of the average number of subhalos as a function of different halo properties for test set $1$, test set $2$ and the small box test set, measured based on the predicted and actual subhalo populations for each test set. Figures \ref{fig:dist1m12}-\ref{fig:dist2m13} show plots of normalized number of subhalos residing in host halos of two different mass ranges as a function of different subhalo properties with subhalos from the predicted and actual populations of test sets $1$ and $2$ (only here, similar figures were not created for the small box test set because there were too few subhalos in the coresponding samples in that set to ensure that their distribution plots were reliable enough for comparisons as one needs to further divide the subhalos in each sample into several bins of their properties in order to create the distribution plots). Figures \ref{fig:xi_set1}-\ref{fig:wp_smallbox} show plots of $3$D and projected correlation functions for various subhalo samples selected from the predicted and actual subhalo populations of test set $1$, test set $2$ and the small box test set. The overwhelming majority of the predictions shown in the testing section have percent errors whose magnitudes are below $5\%$, suggesting a satisfactory performance by the method. 

Finally, in section \ref{sec:discussions}, we started by emphasizing that expanding the set of input halo properties to include parameters beyond what we considered in this paper may improve the performance of the method and can be an interesting topic for a future work. Next, we discussed the application of the method in populating dark matter halos in low resolution N-body simulation boxes with subhalos from a high resolution simulation box and also possible ways to resolve some issues that might arise in such processes. We ended the section by discussing the second application of the method, namely identifying halo properties that are (more strongly) correlated with subhalo abundance/clustering in a given dark matter simulation. 

\section*{Acknowledgements}
We thank Instituto de Astrofisica de Andalucia (IAA-CSIC), Centro de Supercomputacion de Galicia (CESGA) and the Spanish academic and research network (RedIRIS) in Spain for hosting Uchuu DR1, DR2 and DR3 in the Skies $\&$ Universes site for cosmological simulations. The Uchuu simulations were carried out on Aterui II supercomputer at Center for Computational Astrophysics, CfCA, of National Astronomical Observatory of Japan, and the K computer at the RIKEN Advanced Institute for Computational Science. The Uchuu Data Releases efforts have made use of the skun@IAA\_RedIRIS and skun6@IAA computer facilities managed by the IAA-CSIC in Spain (MICINN EU-Feder grant EQC2018-004366-P).

\section*{Data Availability}
All the data used in this work are publicly available in \\ \href{https://www.skiesanduniverses.org/Simulations/Uchuu/}{www.skiesanduniverses.org/Simulations/Uchuu/}. 

\section*{Code Availability}
All code written for the implementation of the method is available upon request.

\bibliographystyle{JHEP}
\bibliography{Refs}

\end{document}